\documentclass[12pt]{article}

\usepackage{latexsym}

\font\tenmsbm=msbm10 scaled 1200
\font\sevenmsbm=msbm9
\newfam\msbmfam
\textfont\msbmfam=\tenmsbm
\scriptfont\msbmfam=\sevenmsbm

\makeatletter
\@addtoreset{equation}{section}
\makeatother

\newcommand{\eref}[1]{(\ref{#1})}

\def\beq{\begin{equation}}
\def\eeq{\end{equation}}
\def\bea{\begin{eqnarray}}
\def\eea{\end{eqnarray}}
\def\bet{\begin{tabular}}
\def\eet{\end{tabular}}
\def\pa{\partial}

\def\ve{\varepsilon}

\def\yh{\hat y}

\catcode`@=11

\begin{document}

\begin{titlepage}

\begin{flushright}
Preprint DFPD 03/TH/03\\
February 2003\\
\end{flushright}

\vspace{0.5truecm}

\begin{center}

{\Large \bf Chern--kernels and anomaly cancellation in $M$--theory}

\vspace{2.5cm}

K. Lechner\footnote{kurt.lechner@pd.infn.it} and  P.A.
Marchetti\footnote{pieralberto.marchetti@pd.infn.it}
\vspace{2cm}

 {
\it Dipartimento di Fisica, Universit\`a degli Studi di Padova,

\smallskip

and

\smallskip

Istituto Nazionale di Fisica Nucleare, Sezione di Padova,

Via F. Marzolo, 8, 35131 Padova, Italia
}

\vspace{2.5cm}

\begin{abstract}
This paper deals with magnetic equations of the type $dH=J$ where the current
$J$ is a $\delta$--function on a brane worldvolume and $H$ a $p$--form
field strength. In many situations in $M$--theory this equation
needs to be solved for $H$ in terms of a potential. A standard universality
class of solutions,
involving Dirac--branes, gives rise to strong intermediate singularities in
$H$ which in many physically relevant cases lead to inconsistencies. In this
paper we present an alternative universality class of solutions for magnetic
equations in
terms of Chern--kernels, and provide relevant applications, among
which the anomaly--free effective action for open $M2$--branes ending on
$M5$--branes.
The unobservability of the Dirac--brane requires a Dirac
quantization condition; we show that the requirement of ``unobservability''
of the Chern--kernel leads in $M$--theory to classical gravitational
anomalies which cancel precisely their quantum counterparts.

\vspace{0.5cm}

\end{abstract}

\end{center}
\end{titlepage}

\newpage

\baselineskip 6 mm


\section{Introduction and Summary}

Extended objects represent basic excitations of $M$--theory and string
theory; usually they show up in pairs of a brane with electric charge $e$
and a dual brane with magnetic charge $g$; the consistency condition for
their coexistence is represented by Dirac's quantization condition
 \beq
 \label{Dirac}
 eg=2\pi n\, G,
 \eeq
where $G$ is Newton's constant and $n$ is an integer. In terms of
the charges the tensions are given by
 \beq
T_e=e/G,\quad T_g=g/G.
 \eeq
The distinction between branes and dual branes is to a certain
extent conventional, we call here (electric) ``branes'' the lower
dimensional objects and (magnetic) ``dual branes'' the higher
dimensional ones. The distinction becomes indeed physical when one
considers the couple of classical Maxwell equations arising from
supergravity theories with electric and magnetic sources: the
electric equation for a ``brane''--source gets non--linear
corrections, while in most cases the magnetic one for a ``dual
brane''--source does not and remains of the form
 \beq
\label{mag}
dH=gJ_g,
 \eeq
where $J_g$ is the $\delta$--function on the dual brane worldvolume
$M_g$ and $H$ is the
field strength curvature form. In the absence of non--linear corrections the
electric Maxwell equation is
 \beq
\label{ele}
d*H=eJ_e,
 \eeq
where $J_e$ is the $\delta$--function on the brane worldvolume $M_e$.
This system of equations
is classically consistent if both brane worldvolumes are closed, i.e. $dJ_e=
0=dJ_g$. If one of the two branes is absent, say the magnetic one, then it is
also
straightforward to write an action because \eref{mag} can be solved locally
in terms of a potential through $H=dB$, but if both branes are present
then the introduction of a potential is more subtle. In this case a standard
device to solve the magnetic equation in terms of a potential requires the
introduction of a
Dirac--brane, i.e. a surface whose boundary is $M_g$, and one can
write an action which is well defined modulo $2\pi$ if \eref{Dirac} holds.
In this case the potential $B$ carries necessarily a singularity along
the Dirac--brane, and therefore also on $M_g$ which is its boundary.
For linear Maxwell equations with sources one can clearly invert the role of
branes and dual branes, upon replacing $H\rightarrow *H$ and introducing a
dual potential $\widetilde B$ instead of $B$.

The present paper deals with situations where the r.h.s. of \eref{ele} carries
non--linear corrections which typically arise in supergravity theories coupled
to branes. Examples are $M5$-- and $M2$--branes in eleven dimensions,
$NS5$--branes and $NS$--strings in ten dimensions, and $D$--branes in type
$II$ theories. First of all, due to these corrections, in this case it is not
possible to introduce a dual potential $\tilde B$ {\it without} introducing
a potential $B$; this means that to write an action one must {\it necessarily}
solve \eref{mag} in terms of $B$. Once this has been done one can regard
the non--linear version of \eref{ele} as equation of motion and try to write
an action for this equation (so eventually there is no need to introduce
$\widetilde B$).

Since the magnetic equation remained the same as in the linear
case a natural attempt to solve it would be again through the
introduction of a Dirac--brane. To see if such a procedure works
we have to take a closer look on the nature of the non--linear
corrections to the r.h.s. of \eref{ele}. In all the examples
quoted above these corrections contain terms of the kind 
 \beq
\label{nonlin} BJ_g\quad  {\rm or} \quad \Phi J_g, 
 \eeq for some
target space form $\Phi$, not involving $B$ \footnote{Clearly the
term $BJ_g$ can not appear alone in the equation of motion since
it would spoil gauge invariance under $\delta B=d\Lambda$. A
complete invariant equation of motion will be given in section
five.}. In the first term the potential $B$ gets evaluated
directly on $M_g$; on the other hand, to get a term like $\Phi
J_g$ in the equation of motion one has to add to the action a term
like $\int B \Phi J_g=\int_{M_g}B\Phi$, and this requires again to
evaluate $B$ on $M_g$. But, as observed above, in the presence of
a Dirac--brane $B$ is singular on $M_g$ and hence its restriction
to (pullback on) $M_g$ is not defined. We conclude that
whenever the dynamics of a theory involves a potential solving
\eref{mag} {\it evaluated on the dual brane worldvolume $M_g$
itself} one has to abandon the Dirac--brane approach for that
brane. This seems a dangerous conclusion because the quantization
condition \eref{Dirac} is intimately related with the use of
Dirac--branes.

The main result of this paper is that a way out is provided by
Chern--kernels \cite{Chern}: these are differential forms $K$, satisfying
$dK=J_g$, generalizing the Coulomb electric field.
The fundamental problem with Dirac--branes is
that they introduce $\delta$--like singularities along their worldvolume;
the forms $K$ instead exhibit milder {\it universal inverse--power--like
singularities on $M_g$}, and they allow to define potentials $B$ which are
well--defined on $M_g$. They constitute a new universality class of
solutions for the magnetic equation $\eref{mag}$ and they allow to write
well--defined effective actions for supergravity theories with electric
and/or magnetic sources.

Since the non--linear terms in \eref{nonlin} are independent of
electric sources, we stress that in the presence of such terms the
Chern--kernel approach is needed to write a well--defined action,
{\it even }if there are only magnetic sources and no electric
ones.

Apart from solving a problem which could look rather formal, in relevant
situations Chern--kernels lead to classical effective actions which exhibit
{\it necessarily} gravitational anomalies  that cancel in string
theory against their quantum counterparts. This can be understood roughly
as follows. Chern--kernels are not uniquely defined since the
universality class admits infinitely many representatives. Two
representatives differ by $K^\prime=K+dQ$ and the classical action
has to be invariant under such shifts; in other words the Chern--kernel
has to be ``unobservable''.
To save this invariance one has
to add counterterms to the classical action which, in turn, give rise
to gravitational anomalies supported on $M_g$. In all
examples we examined these classical anomalies cancel precisely against
their quantum counterparts, produced by the chiral fields localized on
$M_g$.

In principle another way out to solve the above problem, i.e. to
keep $B$ free from singularities on the magnetic brane, would be to
substitute the $\delta$--like current $J_g$ with a regularized
smooth one $\widetilde J_g$, entailing the same total magnetic
charge. This procedure has been used in \cite{Harvey, Becker}
where the gravitational anomaly cancellation for $M5$--branes in 
$D=11$ has been discussed for the first time. A problem related with such
an approach is that at the quantum level a minimal coupling of a smooth 
magnetic current to an elementary electric source 
(i.e. an $M2$--brane with $\delta$--like support) spoils the 
unobservability of the electric Dirac--brane. This point will 
be clarified in the next section. The authors of reference 
\cite{BK} performed a redefinition of the three--form potential $B$ used in 
\cite{Harvey}, which allows to remove the regularization. In this adapted
version, which has been further applied to a five--dimensional
$M5$--brane toy--model in \cite{BHR}, the approach of \cite{Harvey,Becker,BK} 
becomes indeed comparable with the Chern--kernel approach. The relevant 
differences will be discussed at the end of section five. 

Self--dual branes play a special role and exhibit an additional
peculiar feature with respect to a system of branes/dual--branes.
By definition a self--dual brane is a brane minimally coupled to a
self--dual field strength in a $(4N+2)$--dimensional space--time.
This means that even in the absence of non--linear terms the
minimal coupling $\int_M B$ becomes problematic if $B$ has
Dirac--brane--type singularities along $M$. As we will show, in
this case in principle both possibilities  -- a Dirac--brane (with
a framing regularization) or a Chern--kernel -- are allowed, but
they are physically inequivalent: in the Dirac--brane approach the
dynamics is consistent if the charge is quantized, while in the
Chern--kernel approach there is no need of charge quantization,
but a (classical) gravitational anomaly shows up, which for
consistency should be canceled by a quantum counterpart.
Eventually for self--dual branes it is the particular physical situation 
that decides which approach one has to use.

In section two we review briefly the Dirac--brane approach,
indicating to which extent it can be circumvented in compatibility with
\eref{Dirac}. In section three we define Chern--kernels for even
and odd ranks and state their main properties, while
section four is devoted to self--dual branes. In section five we
illustrate how Chern--kernels work in a highly non trivial case:
$M2$--branes ending on $M5$--branes. This is actually a more
general case then the one we referred to above since
the electric current is not conserved and there are quantum
gravitational anomalies not only on the (magnetic) $M5$--brane,
but also on the boundary of $M2$; this boundary corresponds indeed
to a self--dual string. Actually, in this case anomaly
cancellation has been realized until now only partially
\cite{Mourad}. Section six gives a further application: a
$B_3\leftrightarrow \widetilde B_6$--duality symmetric action for
bosonic $D=11$ supergravity with electric and magnetic sources.
Section seven is devoted to concluding remarks.

The present paper provides also some proofs
which, for lack of space, have not been reported in \cite{LMT}.

A comment on our mathematical framework is in order.
Since our currents $J$ carry a $\delta$--like support they are not
smooth differential $p$--forms, they are rather distribution valued
so called $p$--currents \cite{deRham}. 
Consistency then requires that all our differential forms
have to be considered in the space of distributions and that also the
differential operator $d$ acts in the sense of distributions. This implies
that we have {\it always}
$$
d^2=0,
$$
and that in a topologically trivial space--time a closed form is always
exact. 
In this framework two caveats are in order: first, the
product of two distributions is in general not a distribution if the
singularities involved are too strong; second, Leibntz's rule does
in general not hold: for even forms the relation
$d(\phi\psi)=\phi d\psi+d \phi\psi$ may not be valid, even if the product
$\phi\psi$ is well--defined (a trivial example is $\phi=\psi=\varepsilon(x)$,
the sign function). As we will see, in the Chern--kernel approach all our
products will amount to well--defined distributions, and we will have to worry
only about Leibnitz's rule. Henceforth, according to the standard physical 
terminology, we will call the $p$--currents again $p$--forms.  

In most of the paper we assume to be in a topologically trivial space--time,
some topological aspects will be discussed in the concluding section.

\section{Dirac--branes}

We review here briefly the standard Dirac--brane approach, allowing to write
an action for the linear system \eref{mag}, \eref{ele}. $J_g$ is the
$\delta$--function on the closed worldvolume $M_g$, more precisely
it is the ``Poincar\`e--dual in the space of $p$--currents'' of $M_g$: such 
forms are called integer forms.
For a precise definition and basic properties of this Poincar\`e--duality see 
e.g. \cite{deRham,LM}. Here we recall that by definition the Poincar\`e--dual
$J_M$ of a manifold $M$ satisfies
 $$
\int\Phi J_M=\int_M \Phi,
 $$
for every test form $\Phi$. The integral of
products of integer forms, closed or not, is always integer whenever it is
well defined.
The operator $d$ corresponds to the boundary operator $\partial$ on manifolds.
If multiplied by a target--space form, $J_M$ performs the
pullback on $M$:
 $$
\Phi J_M = \Phi|_M J_M.
 $$

Let $M_g$ be a closed $d$--dimensional manifold in a $D$--dimensional
space--time. 
Since $M_g$ is closed we can introduce a $(d+1)$--surface $S_g$  -- a
Dirac--brane -- whose boundary is $M_g$, $M_g=\partial S_g$. The
Dirac--brane is clearly not unique: choosing another Dirac--brane we
have $M_g=\partial S_g^\prime$ and there exists an interpolating
$(d+2)$--surface $T_g$ such that $S_g^\prime-S_g=\partial T_g$. The
corresponding relations for the Poincar\`e--duals of these
surfaces are
 \beq
 \label{transC}
 J_g=dC_g=dC_g^\prime, \quad
C_g^\prime-C_g=dW_g,
 \eeq 
where $J_g$, $C_g$ and $W_g$ are integer forms of rank 
$(D-d)$, $(D-d-1)$ and  $(D-d-2)$ respectively.

The first relation of \eref{transC} permits to solve
\eref{mag} in terms of a potential $B$
 \beq
 \label{H1} H=dB+gC_g,
 \eeq 
and to write then an action giving rise to \eref{ele}:
  \beq
\label{I1} I_1={1\over G}\int\left({1\over 2}H*H-eB
J_e\right)\equiv I_{kin}+I_{wz},
 \eeq where according to string
theory  for the overall normalization we used Newton's constant.

To keep the field strength invariant under a change of
Dirac--brane we have to require that $B$ transforms as 
 \beq
\label{transB} B^\prime= B-gW_g, 
 \eeq 
where, we recall, $W_g$ is
the $\delta$--function on the surface $T_g$. Under this
transformation $I_{kin}$ is manifestly invariant while the
Wess--Zumino term changes as
 \beq
 \label{change1}
 \Delta I_{wz}={eg\over G}\int W_gJ_e.
 \eeq
Since the integrand is a product of integer forms, the
integral is integer and $\Delta I_{wz}$ is an integer multiple of
$2\pi$ thanks to charge quantization \eref{Dirac}. From
\eref{transB} we retrieve that the potential $B$ is ill--defined
on $M_g$ because the form $W_g$ is the $\delta$--function on the
surface $T_g$ which, by construction, contains as submanifold
$M_g$; so $W_g$ does not admit pullback on $M_g$ and due to
\eref{transB} $B$ bears the same fate.

This is in synthesis the standard Dirac--brane construction of an
action for Maxwell equations in the presence of electric and
magnetic sources. Let us now see how stringent it is. First we
note that one can introduce also an {\it arbitrary} antiderivative
$K_e$ for the conserved electric current, $J_e=dK_e$, and rewrite
the Wess--Zumino term as
 \beq \label{trucco} I_{wz}={e\over G}\int
dB K_e={e\over G}\int HK_e -{eg\over G}\int C_gK_e.
 \eeq So in the
Dirac--brane approach, as it stands, one has really an asymmetric
treatment for electric and magnetic sources: for the magnetic
source one introduces a Dirac--brane as antiderivative, while for
the electric one one can introduce an arbitrary antiderivative;
but due to duality one expects that the situation can be reversed.
This is indeed the case. Choose for $K_e$, which is arbitrary, the
Poincar\`e--dual $C_e$ of an electric Dirac--brane, $J_e=dC_e$.
Then the integrand in the last term in \eref{trucco} is a product
of integer forms, and the term itself becomes an integer multiple
of $2\pi$ due to charge quantization; it can therefore be
disregarded. The action $I_1$ can then be rewritten as
 \beq \label{I2}
I_2={1\over G}\int\left({1\over 2}H*H+eHC_e\right),
 \eeq where the
magnetic Dirac--brane does now no longer appear explicitly. This
means that in this form of the action one can take
 \beq
\label{H2}
H=dB+gK_g,
 \eeq with $K_g$ an {\it arbitrary} antiderivative of
the magnetic current,
 \beq \label{anti}
 J_g=dK_g.
 \eeq As $I_1$
had a spurious dependence on the magnetic Dirac--brane, the action
$I_2$ has now a spurious dependence on the electric Dirac--brane;
for $C_e^\prime=C_e+dW_e$, with $W_e$ the $\delta$--function on a
suitable surface, $I_2$ changes by
 \beq\label{change2}
 \Delta I_2={e\over G}\int HdW_e=-{eg\over G}\int J_gW_e,
 \eeq
again an integer multiple of $2\pi$. It is obvious that $I_2$
gives as equation of motion still \eref{ele}. So the situation is
now indeed reversed: with the price of introducing an
(unobservable) electric Dirac--brane, we can choose for the
magnetic brane an arbitrary form $K_g$ satisfying \eref{anti}.

With this simple observation we have now a new possibility for
introducing a potential. The main difference between \eref{H1} and
\eref{H2} is the following. We have $dK_g=J_g=dC_g$, and this means that
$C_g$ as well as $K_g$ are necessarily singular on the magnetic brane because
$J_g$ is the $\delta$--function on $M_g$.
$C_g$ has $\delta$--function like singularities along a Dirac--brane, say
along a fixed space--time direction; these singularities are not
universal since there is no preferred direction in space--time:
changing the Dirac--brane changes this space--time direction and
hence the support of the singularities, but this means that $B$
itself has to change by singular terms -- see \eref{transB} -- interpolating 
between the two different singular behaviours, see also 
\cite{LM,LM2}. Consider now
the second possibility \eref{H2}. In this case $K_g$ has to
satisfy a priori only  $dK_g=J_g$ with no further
restrictions; but then, as we know from the Coulomb
electric field whose divergence is a $\delta$--function, its
singularities can be also of the inverse--power--type, i.e. 
milder then $\delta$--type singularities: this behaviour is indeed
realized by a Chern--kernel. The problem with
inverse--power--like singularities is that, as in the Dirac--brane
case, a priori they can still change when $K_g$ changes by a
closed form. This problem will be solved in the next
section.

From \eref{change1} and \eref{change2} it is also clear that
magnetic and electric branes can be consistently minimally coupled
at the quantum level only if {\it both
currents} carry a $\delta$--like support i.e. if both are integer
forms, otherwise either $W_e$ ($W_g$) or $J_g$ ($J_e$)  would not
be an integer form and the integrals appearing in \eref{change1},
\eref{change2} would not be integer: as consequence under a change of 
Dirac--brane the action
would not change by an integer multiple of $2\pi$, even if the
charges are quantized. 

\section{Chern--kernels}

Chern--kernels are differential forms $K$ satisfying
 \beq
\label{solita}
J=dK,
 \eeq
with a specific singular behaviour near the brane. We take $J$
as an $n$--form whose Poincar\`e--dual is a closed
$(D-n)$--dimensional brane worldvolume $M$ in a $D$--dimensional
space--time; the Chern--kernel is then an $(n-1)$--form. The main
motivation for an analysis of its properties is that its defining
relation allows to solve the magnetic equation $dH=gJ$ as
 \beq
 \label{H22}
 H=dB+gK.
 \eeq

\subsection{Brane geometry}

We introduce first the main geometrical quantities which are
defined on the brane worldvolume $M$. The brane is parametrized
locally by $x^\mu(\sigma)$, $\mu=(0,\cdots,D-1)$, and carries a
$(D-n)$--dimensional tangent space spanned by
$E_i^\mu(\sigma)\equiv\pa_i x^\mu(\sigma)$, $(i=0,\cdots,D-n-1)$.
The normal $SO(n)$--fiber is spanned by the vectors
$N_\mu^a(\sigma)$ satisfying
 \beq\label{normal}
N_\mu^a E^\mu_i=0,\quad N_\mu^a
N^{\mu b}=\delta^{ab},
 \eeq
where $a=(1,\cdots,n)$. These basis
vectors for the normal space are defined modulo a local
$SO(n)$--rotation,
 \beq
 N_\mu^{\prime a}(\sigma)=\Lambda^{ab}(\sigma)N_\mu^b(\sigma).
 \eeq
On $M$ we can introduce also an $SO(n)$--connection
$a^{ab}(\sigma)$ \footnote{This connection can be parametrized in
terms of the normal vectors by $a^{ab}=N^{\mu
b}\left(dN_\mu^a+\Gamma_\mu^\nu N_\nu^a\right)$, where
$\Gamma_\mu^\nu$ is the pullback on $M$ of the affine space--time
connection. We suppose to work in a curved space--time, so the
indices $\mu,\nu,\ldots$ are raised and lowered with the metric
$g_{\mu\nu}$ evaluated, in case, on $M$.}, and its curvature
$f^{ab}=da^{ab}+a^{ac}a^{cb}$. For later purposes it is convenient
to extend $a$ to an $SO(n)$--connection $A(x)$ on the whole
space--time, that is asymptotically flat along directions
orthogonal to the brane. We call the corresponding curvature
$F\equiv dA+AA$,
 \beq
 \label{restr}
A|_M=a,\quad F|_M=f,
 \eeq
where $|_M$ means pullback of a space--time form on $M$. While $a$
and $f$ are physical data, the quantities $A$ and $F$ are not: so
when they are used at intermediate steps, eventually one has to
show that physics is independent of these unphysical data. This means that 
one can look at $A$ and $F$ as ``holographic extensions'' of $a$ and $f$
from $M$ to the whole space--time, which do not introduce new degrees of 
freedom.

To $M$ we can associate also systems of normal coordinates. Such a
system realizes a diffeomorphism from the coordinates $x^\mu$ to
the coordinates $(\sigma^i,y^a)$, with $i=0,\cdots,D-n-1$ and
$a=1,\cdots,n$, specified by the functions $x^\mu(\sigma,y)$. As a
power series in $y$ -- the normal coordinates -- one has
 \beq
 \label{series}
x^\mu(\sigma,y)=x^\mu(\sigma)+y^aN^{\mu a}(\sigma)+o(y^2).
 \eeq
Inversion of the diffeomorphism leads to the space--time functions
$y^a(x)$ with
 \beq
\label{defy}
 y^a|_M=0, \quad  \pa_\mu y^a|_M=N_\mu^a.
 \eeq
Throughout this paper we suppose that the functions $y^a(x)$, with
this behaviour on $M$, are globally defined; for more general
situations see \cite{LMT} and section seven. Again, the physical content 
of these
functions is only their behaviour \eref{defy} on $M$, their values 
away from $M$ correspond to unphysical data which
eventually have to be unobservable.

Actually, the behaviour \eref{defy} on $M$ is defined up to an
$SO(n)$--transformation, as are the normal coordinates themselves.
Formally, through the functions $A^{ab}(x)$ and $y^a(x)$, we have
thus extended the $SO(n)$--structure to the whole space--time. In
particular we can define the $SO(n)$--covariant derivatives
$Dy^a=dy^a+y^bA^{ba}$.

For future purposes it is convenient to introduce an additional
(equivalent) $SO(n)$--connection ${\cal A}$ and related curvature
${\cal F}$. Set
$$
\yh^a=y^a/y,\quad \yh^a\yh^a=1,
$$
and define
 \beq\label{shift}
{\cal A}^{ab}=A^{ab}-2\,\yh^{[a}D\yh^{b]} \equiv A^{ab}+X^{ab}.
 \eeq
This gives for the curvature
 \beq\label{curv}
{\cal F}^{ab}=d{\cal A}^{ab}+{\cal A}^{ac}{\cal A}^{cb}=
F^{ab}+D\yh^aD\yh^b+2\yh^{[a}F^{b]c}\yh^c.
 \eeq
The qualifying properties of this connection are that its
curvature has vanishing components along $\yh^a$ and that $\yh^a$
is covariantly constant with respect to it,
 \beq
\label{project} \yh^a {\cal F}^{ab}=0, \quad D({\cal
A})\yh^a\equiv d\yh^a+\yh^b{\cal A}^{ba}=0.
 \eeq
Notice that, contrary to $A$ and $F$, the forms ${\cal A}$ and
${\cal F}$ do not admit pullback on $M$.

\subsection{The Coulomb form}

In terms of normal coordinates the current can be expressed in an
$SO(n)$--invariant way as
$$
J={1\over n!}\,\ve^{a_1\cdots a_n}dy^{a_1}\cdots
dy^{a_n}\delta^n(y),
$$
signaling of course that the brane stays at $y=0$.

In searching for an $(n-1)$--form whose differential equals $J$ a
first attempt would be to consider a generalization of the
three--dimensional Coulomb electric field, more precisely of its
Hodge--dual
$$
K_0={1\over 8\pi}\,\ve^{abc}{1\over y^3}\,y^ady^bdy^c= {1\over
8\pi}\,\ve^{abc}\yh^ad\yh^bd\yh^c,
$$
corresponding to $n=3$. This
formula can also be regarded as the angular form on a two--sphere,
with unit integral. Its generalization for a generic $n$ is:
 \beq
\label{coulomb} K_0={(-)^{n+1}\,\Gamma\left(n/2\right) \over 2
\pi^{n/2}(n-1)!}\,\ve^{a_1\cdots a_n}\yh^{a_1}d\yh^{a_2}\cdots
d\yh^{a_n},
 \eeq
and it is easy to see that indeed
 \beq
 \label{ancora}
dK_0=J.
 \eeq
The problem with the Coulomb form in \eref{coulomb} is that its
(singular) behaviour on $M$, i.e. as $y\rightarrow 0$, is not
universal but depends on the particular normal coordinate
functions $y^a(x)$ we have chosen. In other words, it is invariant
under global $SO(n)$--rotations, but not under local ones: for
different normal coordinates one has indeed $\yh^{\prime
a}=\Lambda^{ab}\yh^b$, where $\Lambda^{ab}(x)$ is an
$SO(n)$--matrix. For the Coulomb form associated to the rotated
normal coordinates we have of course still $dK_0^\prime=J$, and
hence
$$
K_0^\prime - K_0 = dQ_0,
$$
but the form {\it $Q_0$ is singular on $M$}, because the matrix
$\Lambda^{ab}(x)$ does not reduce to the identity on $M$, meaning
that $K_0$ and $K_0^\prime$ exhibit different singularities.

The question is then if $K_0$ admits a completion $K$ carrying an
{\it $SO(n)$--invariant singular behaviour} on $M$, which solves
still the magnetic equation \eref{ancora}. Clearly one should then
have
 \beq\label{dec}
K=K_0+d\Phi,
 \eeq
for a convenient form $\Phi$. The answer to this question is
affirmative, but since it entails completely different features
for even and odd currents we treat the two cases separately. The
difference originates essentially from the fact that the Euler
characteristics of an odd bundle is zero.

\subsection{Even Chern--kernels}

For $n$ odd the Coulomb--form can be completed to an $SO(n)$--invariant
Chern--kernel given by \cite{charact}
 \beq
\label{even} K= {\Gamma\left(n/2\right) \over 2
\pi^{n/2}(n-1)!}\,\ve^{a_1\cdots a_n}\,\yh^{a_1}\,{\cal
F}^{a_2a_3}\cdots {\cal F}^{a_{n-1}a_n},
 \eeq
where ${\cal F}^{ab}$, see equation \eref{curv}, reduces here
actually to $F^{ab}+D\yh^aD\yh^b$, due to the presence of the
factor $\yh^{a_1}$.
$SO(n)$--invariance is manifest, and one has only to show that
$dK=J$. Consider first the $\delta$--function contribution to
$dK$. It is immediately seen that $K$ contains as building block
$K_0$ which, for dimensional reasons, is the unique term which can
give rise to a distribution--valued contribution, and we know
already that $dK_0=J$. It remains to show that formally, i.e.
neglecting the $\delta$--function contribution, $K$ is closed.
This is obvious if one observes that, since $K$ is invariant, the
differential $d$ can be substituted with the differential
covariant w.r.t. ${\cal A}$, $dK=D({\cal A})K$. The conclusion follows 
then from the second relation in \eref{project} and from $D({\cal A}){\cal
F}=0$.

Since the expression for $K$ is a polynomial function of $\yh$ and
$A$ it is also clear that \eref{dec} holds with $\Phi$ polynomial
in $\yh$ and $A$ as well. For $n=3$ one has e.g.
 $$
K={1\over 8\pi}\,\ve^{abc}\yh^a \left(F^{bc}+D\yh^bD\yh^c\right),
\quad
 \Phi={1\over 8\pi}\,\ve^{abc}\yh^a A^{bc}.
 $$
 With respect to $K_0$, which carries $n$ powers of (the
singular functions) $\yh$, the even Chern--kernel contains
additional subleading singular terms with powers of $\yh$ ranging
from 1 to $n-2$: these subleading terms are required to form an
invariant singular behaviour.

As it stands $K$ depends on the data $A^{ab}(x)$ and $y^a(x)$,
whose behaviour {\it away} from $M$ is unphysical. The dependence
on these data is in some sense analogous to the dependence on the
Dirac--brane in the Dirac--brane approach, and it must be
compensated by a transformation of the potential $B$, see
\eref{H22}. Changes in these data $A\rightarrow A^\prime$,
$y\rightarrow y^\prime$ are constrained  by the boundary values on
$M$, \eref{restr} and \eref{defy}, where {\it they must reduce to an
$SO(n)$--rotation}. Since we have in any case
$\yh^{\prime a}\yh^{\prime a}=1$, the most general changes can be
parametrized by
 \beq
 \label{changes}
 \yh^\prime = \Lambda \yh,\quad A^\prime= \Lambda (A+W) \Lambda^T
 -\Lambda d\Lambda^T,
 \eeq
where $\Lambda(x)\in SO(n)$ and the arbitrary Lie--algebra valued one--form
$W$ is constrained only to vanish on $M$,
 \beq
 \label{change}
W|_M=0.
  \eeq
Consider now the relation between two different Chern--kernels;
since the differential of a Chern--kernel equals always $J$ it is
clear that the difference is an exact form,
 \beq\label{qt1}
K^\prime =K+dQ.
 \eeq
But \eref{change} implies that moreover
 \beq
 \label{qm}
 Q|_M=0.
 \eeq
Indeed, due to $SO(n)$--invariance
 $$
 K^\prime -K=K(A^\prime,\yh^\prime)-K(A,\yh)=
 K(A+W,\yh)-K(A,\yh),
 $$
and since $K_0$ is independent of $A$ \eref{dec} gives
 $$
Q=\Phi(A+W,\yh)-\Phi(A,\yh),
 $$
 which contains at least one power of $W$. For $n=3$ one has
 $$
Q={1\over 8\pi}\,\ve^{abc}\yh^a W^{bc}.
 $$
Equation \eref{change} implies then \eref{qm}.

This shows not only that the even Chern--kernel has a universal
singular behaviour on $M$ (for that it would be sufficient to show
that $Q|_M$ is a well defined form), but also that the pullback of
the  potential $B$ is a completely invariant form: since
invariance of the field strength in \eref{H22} demands
 \beq
 \label{qt2}
B^\prime=B-gQ,
 \eeq
one gets
 $$
 B^\prime|_M=B|_M.
 $$
We call the transformations \eref{qt1}, \eref{qt2} {\it
$Q$--transformations}; the dynamics of a physical theory must be
independent of the particular Chern--kernel used as antiderivative
of $J$, meaning that its action has to be invariant under such
transformations. $Q$--invariance will therefore be a guiding
principle for constructing consistent actions for magnetic branes
with non--linear interactions; in the presence of closed electric branes
we must, in addition, require independence of the action of the
electric Dirac--brane. In this sense $Q$--invariance plays a role
similar to unobservability of a Dirac--brane: the transformations
\eref{transC} and \eref{transB} coincide indeed formally with
\eref{qt1} and \eref{qt2}.

We emphasize that in this framework the physical gravitational normal bundle
$SO(n)$--transformations on $M$ arise as $Q$--transformations restricted to
$M$, with transformation parameter $\lambda=\Lambda|_M$, see \eref{changes}.

Since even Chern--kernels have an invariant behaviour near the
brane and lead, therefore, to potentials which are invariant when
evaluated on the brane, one might think that even Chern--kernels
have nothing to do with gravitational anomalies localized on the
brane; this is however not the case since products of an even
number of Chern--kernels are cohomologically equivalent to
characteristic classes, i.e. to invariant polynomials of the
normal bundle $SO(n)$--curvature $F$ which, upon descent, give rise to
gravitational anomalies. An important example in which the even
Chern--kernel--approach leads to cancellation of the quantum
normal bundle anomaly is represented by the $M5$--brane, see
\cite{LMT}. For this reason we discuss in the next subsection the
basic properties of such products.

\subsection{Products of even Chern--kernels}

In this section we derive algebraic relations involving
powers of the even Chern--kernel, which are needed in the
construction of effective actions for $p$--branes in $M$--theory,
see section five.

In general the product of two distributions does not define a
distribution, but in the case of even Chern--kernels the product
$K...K$ defines still a distribution--valued differential form,
with again inverse--power--like singularities on $M$. Contrary to
$K$, however, the even powers such as $KK$ are closed forms. This
is not in contrast with $dK=J$ since what fails here is Leibnitz's
rule: the formal computation $d(KK)=2KJ$ makes indeed no sense
because $K$ does not admit pullback on $M$; we must first evaluate
the product and then take the differential. We will actually now
show that there exists a form $S$, polynomial in $A$ and $\yh$ but
not $SO(n)$--invariant, such that (see also \cite{BottCatt,Intri})
 \beq\label{product}
KK={1\over 4}\,d S.
 \eeq
We begin with the evaluation of $KK$. Recalling that $\yh^a {\cal
F}^{ab}=0$, from the definition \eref{even} this product is easily seen to
reduce to a combination of traces of ${\cal F}$. Setting $n=2m+1$ one obtains
 $$
 KK={1\over 4}\,P({\cal F}),
 $$
where $P$ is the $m$--th Pontrjagin form. Remembering
\eref{shift}, a standard transgression formula gives then
 \beq\label{kk}
 P({\cal F})=P(F)+dY,
 \eeq
where $Y$ is an $SO(n)$--{\it invariant} $(2n-3)$--form, polynomial in
$\yh,D\yh$ and $F$, singular on $M$:
 \beq
 \label{y}
 Y=2m\int_0^1 P(F_t,\cdots,F_t,X)\,dt,
 \eeq
where $A_t=A+tX$ and $F_t=dA_t+A_tA_t$. This leads to
\footnote{$S$ differs from the canonical
Chern--Simons form $P^0({\cal A})$, associated to
$P({\cal F})=dP^0({\cal A})$, by a closed form.}
\beq
 \label{P7}
 S=P^0(A)+Y,
 \eeq
where $P^0(A)$ is the Chern--Simons form associated to $P(F)$.
For generic characteristic classes we use the descent notation
 $$
dP=0,\quad P=dP^0, \quad \delta P^0=dP^1.
 $$
The form $S$ is made out of two contributions: $P^0(A)$ is non--invariant but
regular on $M$, while $Y$ is invariant but singular on $M$, due to the
presence of the singular form $X$. For $n=3$ one has
 \bea\nonumber
 P(F)&=&-{1\over 2(2\pi)^2}\,tr F^2,\\
 P^0(A)&=&-{1\over 2(2\pi)^2}\,tr\left(AdA+{2\over 3}A^3\right),\nonumber\\
 Y&=&-{1\over(2\pi)^2}\,\yh^a D\yh^b F^{ab}.\nonumber
 \eea
From the formulae above it is clear that $KK$ is a closed form. What
happens is essentially that $K$ contains only {\it odd} powers of
$\yh$'s -- in particular the Coulomb form whose differential gives
rise to $J$ -- while $KK$ contains only {\it even} powers of
$\yh$'s and so no $\delta$--function contributions can show up in
its differential.

In considering the triple product $KKK$ one encounters the form
$dYK$ which can be seen to be closed too, more precisely one has
 \beq\label{serve}
d(YK)=dYK.
 \eeq
The key point is again to see if the product $YK$ contains terms
whose differential can give rise to $\delta$--function
contributions \footnote{For an alternative proof see appendix A.}.
Such terms must contain precisely $n-1$ powers of
$d\yh$, as $K_0$. In this case we observe that $K$ ($Y$), being
invariant, contains only even (odd) powers of $D\yh$; this means
that $YK$ contains only odd powers of $D\yh$, with maximum power
$n-2$ (because the product of $n$ or more of them is zero by
antisymmetry). So the maximum power of $d\yh$ showing up in $YK$
is $n-2$, and no Coulomb form can appear. This means that one can
compute the differential of $YK$ algebraically, i.e. ignoring
$\delta$--function contributions, and \eref{serve} follows.

Since $P(F)$ is regular on $M$ we can summarize these properties
as
 \beq\label{chain}
 dK=J,\,\quad d(KK)=0,\quad d(KKK)={1\over 4}P(F)J.
 \eeq

\subsection{Odd Chern--kernels}

In this case the current $J$ is an even form, $n=2m$, and a
construction like \eref{even} is no longer available. On the other
hand, the normal bundle of the brane is now even and one can
define an $SO(n)$--Euler form $\chi(F)$, and the associated
Chern--Simons form $\chi^0(A)$, in a standard way:
 \beq\label{euler}
\chi(F)={1\over m! (4\pi)^m}\,\ve^{a_1\cdots a_n}F^{a_1a_2}\cdots
 F^{a_{n-1}a_n}\equiv d\chi^0(A).
 \eeq
We will now show that the odd Chern--kernel can be written as a
sum \footnote{Strictly speaking, the ``odd spherical kernel'' introduced by
Chern \cite{Chern} is $\Omega$.}

 \beq\label{odd}
 K=\Omega +\chi^0(A),
 \eeq
where $\Omega$ is the $SO(n)$--invariant $(n-1)$--form, polynomial in $\yh,
D\yh$ and $F$, given in \eref{trans}. So the main difference between even 
and odd Chern--kernels is that the former are $SO(n)$--invariant forms,
while the latter are not, due to the presence of the
Chern--Simons form.

To prove that the expression in \eref{odd} satisfies $dK=J$, and
to find an explicit expression for $\Omega$ we start from the
observation that the Euler form of the curvature ${\cal
F}$ vanishes identically,
 \beq\label{zero}
 \chi({\cal F})=0.
 \eeq
This is a direct consequence of $\yh^a{\cal F}^{ab}=0$
\footnote{More concretely, take the identity $\yh^{[b}{\cal
F}^{a_1a_2}\cdots {\cal F}^{a_{n-1}a_n ]}=0$, which holds because
one has $n+1$ antisymmetric indices, and contract it with
$\yh^b$.}. On the other hand, the shift--relation \eref{shift}
allows to express the Euler form for ${\cal F}$ in terms of the
Euler form for $F$,
 \beq\label{basic}
\chi({\cal F})=\chi(F)+d\Omega-J,
 \eeq
where, defining as above $A_t=A+tX$ and $F_t=dA_t+A_tA_t$,
according to standard transgression one has
 \beq\label{trans}
 \Omega=m\int_0^1 \chi(F_t,\cdots,F_t,X)\,dt.
 \eeq
This explains the algebraic contributions in \eref{basic}; the
subtraction of $J$ is necessary because by direct inspection one
can see that $\Omega$ contains as top form in $d\yh$ precisely the Coulomb 
form $K_0$, and
since our differential acts in the sense of distributions this
implies that $d\Omega$ contains as $\delta$--function contribution
precisely $J$, which has to be subtracted. The  identities
\eref{zero} and \eref{basic} ensure then that $dK=J$.

The evaluation of the transgression formula \eref{trans} is
straightforward, for $n=4$ one obtains e.g.
 \bea
\Omega &=&- {1\over 2(4\pi)^2}\,\ve^{a_1\dots
a_4}\yh^{a_1}D\yh^{a_2} \left( 4F^{a_3a_4}+
\frac{8}{3}D\yh^{a_3}D\yh^{a_4}\right),\label{omega}\\
\chi^0(A)&=&{1\over 2(4\pi )^2}\,\ve^{a_1\dots a_4}\left(
A^{a_1a_2}dA^{a_3a_4}+ {2\over 3}\,A^{a_1a_2}\left(
AA\right)^{a_3a_4} \right),\label{chernsimons}\\
\Phi&=& -{1\over 2(4\pi )^2}\,\ve^{a_1\dots a_4} \yh^{a_1}
    \left( 4d\yh^{a_2} +2\yh^b A^{ba_2}\right) A^{a_3a_4},
 \eea
where the form $\Phi$ refers to the decomposition \eref{dec} which
holds clearly also for odd kernels. Notice in particular in
$\Omega$ the presence of the Coulomb form $K_0$.

Also the odd kernel is subject to $Q$--transformations, i.e. to
the changes $y \rightarrow y^\prime$, $A\rightarrow A^\prime$ as
the even one (see \eref{changes}) and we have also here
$$
K^\prime=K+dQ, \quad B^\prime=B-gQ,
$$
for some $Q$. This time one gets
$$
K(A^\prime,\yh^\prime)-K(A,\yh)=K(A+W,\yh)-K(A,\yh)+
\chi^0(A^\prime)-\chi^0(A+W).
$$
Since the Chern--Simons forms differ by the $SO(n)$--rotation $\Lambda$
we have
$$
\chi^0(A^\prime)-\chi^0(A+W)=d\chi^1(A+W),
$$
leading to
$$
Q=\Phi(A+W,\yh)-\Phi(A,\yh)+\chi^1(A+W),
$$
where in $\chi^1$ we suppressed the dependence on $\Lambda$. The
pullback of $Q$ on $M$ is then again finite, but now different
from zero. Since $W|_M=0$ one gets
$$
Q|_M=\chi^1(a),
$$
where, we recall, $a$ is the (physical) $SO(n)$--connection on
$M$, $a=A|_M$.  For an infinitesimal transformation the form
$\chi^1(a)$ is really the descent of the Euler form on $M$,
$$
\chi(f)=d\chi^0(a),\quad
\delta \chi^0(a)=d\chi^1(a).
$$
For the pullback of the potential we obtain then the anomalous
transformation law
 \beq
B^\prime|_M=B|_M-g\,\chi^1(a).
 \eeq

We can summarize the results of this section as follows. The odd
Chern--kernel is made out of two terms, one is singular on $M$ and
the other is regular. The singular contribution $\Omega$ is
invariant and encodes, therefore, the  singularities of  $K$ in a
universal way. Vice versa, the regular contribution $\chi^0(A)$
transforms non trivially under $SO(n)$. For the potential $B$ this
implies that it admits a finite pullback $B|_M$ that under
$Q$--transformations undergoes an anomalous
$SO(n)$--transformation.

It is clear that the anomalous transformation law for $B|_M$ plays
a role in gravitational anomaly cancellation in $M$--theory; for a
basic examplification of this feature -- for $n=4$ -- in the case
of the anomalies of the $NS5$--brane in $D=10$,
$IIA$--supergravity, see reference \cite{CL}. In the next section
we illustrate the occurrence of odd Chern--kernels in a further
important case: self--dual branes.

For computational reasons sometimes it may be useful, though not
strictly necessary, to have at ones disposal regularized currents
and associated regularized Chern--kernels, i.e. a  family of forms
$K^\ve$ and $J^\ve$ with $J^\ve=dK^\ve$, which are regular at $M$
for any $\ve>0$, such that in the sense of distributions
$$
\lim_{\ve\rightarrow 0}J^\ve = J,\quad \lim_{\ve\rightarrow 0}K^\ve = K.
$$
In appendix A we present a particular
class of such regularizations, which goes under the name of
``real algebraic approximation mode'' \cite{charact}, that keeps the currents
and the even Chern--kernels invariant, and that preserves the transformation
properties of the odd kernels. Such regularized objects are
useful for example in determining the singularity structures of products
involving Chern--kernels and currents.

\section {Self--dual branes and chiral bosons}

Self--dual branes are closed branes with a $2N$--dimensional
worldvolume $M$, coupling in a $(4N+2)$--dimensional space--time
minimally to a chiral boson. More precisely we have the
equations
 \bea
dH&=&gJ\label{magn}\\
H&=&*H
\label{chiral},
 \eea
where the current $J$ is a $(2N+2)$--form, the $\delta$--function on $M$.
The magnetic and electric Maxwell equations are thus identified.

In writing an
action for this system  one has to face two problems; the first regards
the construction of a covariant action for the chiral boson, a problem
which is elegantly solved by the PST--approach \cite{PST}. The second
problem
regards the introduction of a potential $B$. Formally the PST--approach
furnishes the action
 \beq
\label{PST}
  I={1\over 2G}\int\left(H*H-{\cal H}_-*{\cal H}_-\right)-{g\over G}\int_M B,
 \eeq
where ${\cal H}_-=i_v(H-*H)$, and $v$ is the auxiliary
non--propagating PST--vector. This action leads to the equation of
motion $H=*H$, once one has solved the magnetic equation
\eref{magn} in terms of a potential. In principle we have now two
ways for doing so.

{\bf 1) Dirac--branes.} We can introduce a Dirac--brane for $M$,
with Poincar\'e--dual $C$,  and write $J=dC$. This gives
 $$
H=dB+gC,
 $$
and $B$ has the known singularities on $M$. Consequently the
Wess--Zumino term in the action $\int_M B=\int BJ$ is ill--defined, even
in this elementary case with only a minimal interaction. But since
formally $I$ gives rise to the correct $B$--equation of motion it may
nevertheless be meaningful to compute its Dirac--anomaly, i.e. its
response under a change of Dirac--brane. One has $C^\prime=C+dW$,
$B^\prime= B-gW$ where (see \eref{transC}) $W$ is the $\delta$--function
on a manifold $T$ whose boundary is made out of the old and new
Dirac--branes. $I$ changes then by
 $$
 \Delta I={g^2\over G}\int WJ,
 $$
where the integral would count the number of intersections between
$M$ and $T$. But since $M$ is a submanifold of $T$  this integral
is ill--defined.

The situation can be saved by introducing a framing
regularization. We replace the surface $M$ in  $\int_M B$ with a
surface $M_\ve$ (and the current $J$ with $J_\ve$) obtained from
$M$ by displacing each point of $M$ by an amount $\ve$ in an
arbitrary direction. This gives instead of $\Delta I$
 $$
\Delta I_\ve={g^2\over G}\int WJ_\ve,
 $$
where the integral is now well--defined and integer. The
Dirac--brane is then unobservable if the charge is quantized as
$$
{g^2\over G}=2\pi n.
$$

One may ask if this simple regularization procedure could be
invoked also to regularize the self--interactions \eref{nonlin} of
a magnetic brane. This is not the case, for two reasons: first,
the terms in \eref{nonlin} are non--linear (quadratic) while the
equation of motion for a chiral boson ($H=*H$) is linear and, second, the
equation of motion itself is well--defined,
while the non--linear terms in \eref{nonlin} involve $B|_M$ and
are ill--defined.

{\bf 2) Chern--kernels.} We can introduce an odd Chern--kernel
such that $J=dK$ and write
 $$
 H=dB+gK.
 $$
In this case the potential is regular on $M$, the integral $\int_M
B$ is well--defined and we have only to worry about
$Q$--invariance. As we know, the pullback of $B$ changes according
to an $SO(2N+2)$--transformation, $\delta B|_M= -g\chi^1(a)$, and
this implies that $I$ carries the normal bundle gravitational
anomaly
$$
\delta I={g^2\over G}\int_M\chi^1(a),
$$
corresponding to the anomaly polynomial ${g^2\over G }\,\chi(f)$,
i.e. to the $SO(2N+2)$--Euler form. A consistent
dynamics requires then the cancellation of this classical anomaly,
for example by the quantum anomaly produced by chiral fields
localized on $M$.

In conclusion, for self--dual branes both possibilities --
Dirac--branes and Chern--kernels -- are available, and which one
has to be used depends on the physical content of the theory. In
the absence of quantum gravitational anomalies one would use
Dirac--branes together with a framing regularization and impose charge 
quantization, while in their
presence Chern--kernels can play a central role in their
cancellation upon choosing particular values for the charges. The
availability of this second possibility is indeed crucial for the
consistency, at the quantum level, of the situation considered in
the next section.

\section{$M2$--branes ending on $M5$--branes}

The effective action for closed $M5$--branes interacting with
closed $M2$--branes through eleven--dimensional supergravity  has
been constructed in \cite{LMT}; it employs a four--form
Chern--kernel for the $M5$--brane current and realizes the
cancellation of the residual $M5$--brane normal bundle anomaly, which is an
$SO(5)$--anomaly. The $M2$--brane carries an odd--dimensional
worldvolume and entails no anomalies; it is moreover closed, so
the coupling to the $M5$--brane could be performed in a standard
way introducing a Dirac--three--brane, as explained in section two
of the present paper, and the dynamics is quantum mechanically
consistent if charge quantization holds.

But in eleven--dimensional space--time an $M2$--brane can have
also an open worldvolume, under the condition that its boundary
belongs to an $M5$--brane \cite{m2m5}; in this respect
$M5$--branes can be really considered as Dirichlet--branes for
eleven--dimensional membranes, as pointed out in \cite{dbranes}.
The principal differences w.r.t to closed $M2$--branes are the
following; first, for open membranes no natural Dirac--brane can
be defined (and no one is needed); second, the boundary of the
membrane is a string describing a two--dimensional worldvolume and
as such it is plagued by gravitational anomalies \cite{Mourad};
third, since open membranes must end on 5--branes the interaction
between the two surfaces is more intricate then in the closed case
where the relative position of the two surfaces is arbitrary; the
open membrane exhibits instead a contact--interaction with the
5--brane and this leads a priori to additional singularities
located at its boundary. The boundary of the membrane couples,
moreover, minimally to the chiral two--form present on the
5--brane and so it becomes actually a self--dual string.

We will show that all these problematic features are naturally and
elegantly solved by the Chern--kernel approach, if the $M2$--brane
charge $e$ and the $M5$--brane charge $g$ assume their $M$--theory
values
 \beq\label{minimal}
g^3=2\pi G,\quad e=g^2,
 \eeq
leading for the brane tensions to the standard relations
\cite{tensioni1;X8,tensioni2}
 $$
 T_g^3 ={2\pi\over G^2}, \quad T_e=G\,T_g^2.
 $$

The results of the present section are summarized in the
Wess--Zumino action \eref{l12}, based on an odd and an even
Chern--kernel, which cancels the quantum gravitational anomalies
of the system.

\subsection{Normal bundles and quantum gravitational anomalies}

Calling the $M2$--brane worldvolume $M_3$, its boundary $M_2$ and
the $M5$--brane worldvolume $M_6$ we have
 \beq
 \label{ending}
 \pa M_3=M_2 \subset M_6, \quad \pa M_6=0=\pa M_2.
 \eeq

On the $M5$--brane the eleven--dimensional Lorentz group
$SO(1,10)$ reduces to the structure group
 \beq\label{struct}
SO(1,5)\times SO(5),
 \eeq
where $SO(1,5)$ is its tangent group and $SO(5)$ the invariance
group of its normal bundle. Since there are chiral fields
localized at $M_6$, i.e. the chiral two--form potential $b$ and
the 32--component Green--Schwarz spinor $\vartheta$ of $SO(1,10)$,
the structure group is plagued by gravitational anomalies. The
associated $M5$--brane eight--form anomaly polynomial can be
rewritten as the sum \cite{Witten},
 \beq\label{m5anom}
2\pi\left(I_8|_{M_6}+{1\over 24}P_8\right),
 \eeq
where $P_8(f)$ is the second Pontrjagin form of the normal bundle
$SO(5)$--curvature $f$, and $I_8(R)$ is the polynomial of the
target--space $SO(1,10)$--curvature $R$, which corrects eleven--dimensional
supergravity by the term $\int BI_8$ \cite{tensioni1;X8}. Clearly,
since $I_8|_{M_6}$ can be written in terms of normal and tangent
curvatures, the $SO(1,5)$--tangent bundle is anomalous as well.

On $M_2\subset M_6$ the structure group \eref{struct} reduces further
according to
 \bea\nonumber
SO(1,5)&\rightarrow& SO(1,1)\times SO_a(4) \\
SO(5)&\rightarrow & SO_b(4),\nonumber
 \eea
where $SO_a(4)$ is the normal group of $M_2$ with respect to $M_6$, and
$SO(1,1)$ is its tangent group.
The reduction of $SO(5)$ to $SO_b(4)$ occurs along the component normal to
$M_6$ of the {\it unique} direction on
$M_2$ which is normal to $M_2$ and tangent to $M_3$.
In conclusion, the structure group on $M_2$ is
$$
SO(1,1)\times SO_a(4)\times SO_b(4).
$$
The Green--Schwarz spinor reduces on $M_2$ to a set of fermions
which are chiral w.r.t this group and they give therefore
rise to gravitational anomalies; the corresponding four--form
anomaly polynomial has been determined in \cite{Mourad} and reads
\footnote{With respect to reference \cite{Mourad} we added a
factor of 1/2 since the Green--Schwarz fermion is Majorana.}
 \beq\label{m1anom}
\pi\left(\chi_b-\chi_a\right),
 \eeq
where $\chi_{a,b}$ are the Euler--forms of the
$SO_{a,b}(4)$--bundles. Notice in particular that the
$SO(1,1)$--tangent bundle is anomaly free.

A consistent low energy effective action for the system $M2+M5$
must therefore cancel simultaneously the anomaly \eref{m5anom}
supported on $M_6$, and the anomaly \eref{m1anom} supported on
$M_2$.

\subsection{Currents, Chern--kernels and equations of motion}

In this section we have to distinguish target--space forms,
defined in eleven--dimensional space--time, from forms defined
only on $M_6$. Generically we will indicate the former with capital
letters and the latter with small letters. One has to take some care
when considering products of forms which are defined on different
manifolds. Indeed, in general the formal product $\phi_1 \phi_2$
between forms defined on different manifolds does not define a
form, but it defines a form say on $M_2$ if $\phi_2$ is the
$\delta$--function on $M_1$ w.r.t. $M_2$ and $\phi_1$ is a form on
$M_1$, see \cite{LMT,CL}.

We indicate the Poincar\'e--duals of $M_6$, $M_3$ and $M_2$
respectively with $J_5$, $J_8$ and $J_9$, where the subscripts
indicate the degrees of the forms. Since $M_2$ is a submanifold of 
$M_6$ we can also define the Poincar\'e--dual of $M_2$ w.r.t. to $M_6$:
this is a four--form on $M_6$ which we indicate with
$j_4$. Then the relations \eref{ending} translate into
 \beq\label{currents}
 dJ_8=J_9,\quad j_4J_5=J_9,\quad dJ_5=0=dj_4.
 \eeq

The starting point for the construction of the bosonic effective
action for $D=11$ supergravity  interacting with $M2$-- and
$M5$--branes is a set of consistent equations of motion for the
physical fields of the system. These are the space--time metric
and a three--form potential $B$ in the target--space $M_{11}$, the
coordinates and a (Born--Infeld like) chiral two--form $b$ on
$M_6$, and the coordinates on $M_3$. Concentrating on the dynamics
of the potentials $b$ and $B$ and calling their field--strengths
$h$ and $H$ respectively we can write their Bianchi identities
(magnetic equations) and equations of motion as
 \bea\label{1}
dH&=&gJ_5\\
d*H&=&{1\over 2}\,HH+g\,h\,J_5+g^2I_8 +eJ_8\label{2}\\
dh&=&dB|_{M_6}+{e\over g}\,j_4\label{3}\\
h&=&*\,h+{\rm (n.l.t.)},\label{4}
 \eea
where the non--linear terms in the self--duality equation for $h$
\eref{4} correspond to the Born--Infeld lagrangian, see the next section.
The form of these equations is almost uniquely fixed by the
requirement of generalized charge conservation for $H$, i.e. by
the requirement that the r.h.s. of \eref{2} is a closed form
\footnote{Our differential acts from the right: $d(\Phi_m\Phi_n)=
\Phi_md\Phi_n+(-)^nd\Phi_m \Phi_n$.}. However, to check this
closure one must first express $H$ and $h$ in terms of $B$ and
$b$, solving \eref{1} and \eref{3}. According to the strategy
developed in this paper we introduce an even $SO(5)$ four--form
Chern--kernel $K$ for $J_5$, a target--space form, and an odd
$SO_a(4)$ three--form Chern--kernel
$k$ for $j_4$ \footnote{Remember that $SO_a(4)$ is indeed the normal
group of $M_2$ with respect to $M_6$.}, a form on $M_6$:
$$
J_5=dK,\quad j_4=dk.
$$
The $SO(5)$--kernel is written explicitly in \eref{k4} and the
$SO_a(4)$--kernel is just the sum of \eref{omega} and
\eref{chernsimons}. The choice of a Chern--kernel for $j_4$,
instead of a Dirac--brane, is advisable because $M_2$ is a
self--dual string plagued by quantum gravitational anomalies, [see
\eref{m1anom}]. On the other hand, the presence of the products
$hJ_5$ and $HH$ in \eref{2} forbids the use of Dirac--branes
also for $J_5$: the first term contains $BJ_5$ and $dbJ_5$, see
\eref{nonlin}, and the second term, using a Dirac--brane $C_g$
instead of the Chern--kernel $K$, would contain a term $C_gC_g$,
which is the square of a $\delta$--function.

The field strengths read then
 \bea
 H&=&dB+gK\\
 h&=&db+B|_{M_6}+{e\over g}\,k.
 \eea
With these determinations, using \eref{currents} and the product
formulae for even Chern--kernels of section three, it is
straightforward to verify that \eref{2} closes.

Since $k$ is odd, under an $SO_a(4)$--transformation on $M_2$ we have,
as remnant of the $Q$--transformation of $b$, the anomalous
transformation law
 \beq\label{anomb}
 \delta b|_{M_2}=-{e\over g}\,\chi_a^1.
 \eeq

\subsection{Anomaly free effective action}

We write the effective action for our system as
$$
\Gamma={1\over G}\left(S_{kin}+S_{wz}\right) +\Gamma_q,
$$
where $\Gamma_q$ is the quantum effective action carrying the
anomalies \eref{m5anom} and \eref{m1anom}, and the classical
action $S_{kin}+S_{wz}$ must by definition give rise to the
equations of motion \eref{2} and \eref{4}. The kinetic terms are
the standard ones,
 \beq\label{kin}
 S_{kin}=\int_{M_{11}} d^{11}x\sqrt{-g}\,R-{1\over
2}\int_{M_{11}}H*H -g\int_{M_6}d^6\sigma\sqrt{-g}\,{\cal
L}(h)-e \int_{M_3}d^3\sigma\sqrt{-g},
 \eeq
where ${\cal L}(h)$, made out of the Born--Infeld Lagrangian
(written in a manifestly
covariant way a la PST \cite{dbipst}, see subsection 6.1),
contains the kinetic terms for the $M5$--brane coordinates as well
as for $b$. Here it suffices to know that it depends on $b$
only through the invariant field strength $h$.

The Wess--Zumino term is written usually as the integral of an
eleven--form; here we choose the equivalent alternative to write
it as the integral of a closed twelve--from on a
twelve--dimensional manifold with boundary,
$$
S_{wz}=\int_{M_{11}}L_{11}=\int_{M_{12}}L_{12},\quad\quad
L_{12}=dL_{11}, \quad\quad\pa M_{12}=M_{11}.
$$
We are here assuming the absence of topological obstructions. One
advantage of this procedure is the manifest invariance of $L_{12}$
under all relevant symmetries, while usually $L_{11}$ is invariant
only up to a closed form. This approach requires to extend the
target--space fields to ${\bf R}^{12}$, and in particular to
extend every involved $p$--brane to a $(p+1)$--brane whose
restriction on $M_{11}$ reduces to the $p$--brane \cite{LMT}.
Closed branes extend to closed branes, and the boundary of the
extended $M2$--brane has to belong to the extended $M5$--brane.
This procedure maintains then the degrees of the current--forms
$J_5$, $J_8$ etc. and hence of the Chern--kernels and of the
dimensions of the normal bundles. With an abuse of language we
indicate the extended worldvolumes still with their
eleven--dimensional names, e.g. $M_6$, even if it is now a
six--brane with a seven--dimensional worldvolume.

The crucial ingredient of this construction is then the
twelve--form
 \bea\nonumber
L_{12}&=&{1\over 6}\,HHH+{g\over 2}\,h\,dB J_5+g^2I_8H
      +g^3\left(I_7^0 +{1\over 24}\,P_7^0\right)J_5\\
      && +\,e\,HJ_8+{e\over 2}\left(db+B+g\,\chi_b^0\right)J_9,
\label{l12}
 \eea
where we introduced the canonical Chern--Simons forms: $I_8=dI_7^0$,
$P_8=dP_7^0$, $\chi_b=d\chi^0_b$.
This formula is the unique solution to the following three
requirements:
\par\noindent 1) It must give rise to the equations of motion
\eref{2} and \eref{4}.
\par\noindent 2) It must be closed.
\par\noindent 3) It must be $Q$--invariant in the bulk, i.e. apart from
terms supported on the branes.

The equations of motion fix indeed all terms in $L_{12}$ which depend
on $B$ or $b$. The terms which are independent of these fields are
fixed by the requirements 2) and 3).
The formula is, actually, manifestly invariant under $Q$--transformations
of $B$: this potential shows up in the combination $H$ or as $B|_{M_6}$ and
both are invariant. The potential $b$ shows up as  $h$, which is invariant,
and as $dbJ_9$ which transforms by an $SO_a(4)$--transformation according
to \eref{anomb}. Eventually, the terms
$g^3\left(I_7^0 +{1\over 24}\,P_7^0\right)J_5$ and ${ge\over 2}\,\chi_b^0J_9$
are needed to make $L_{12}$ a closed form, as can be seen using
-- before differentiation -- the product formulae of section three
\footnote{That $L_{12}$ is closed for $e=0$,
i.e. in the absence of $M2$--branes, has been shown in \cite{LMT}.}.
The unique non trivial point, in checking that $dL_{12}=0$, is to show that
{\it there exists a three--form $X$ on $M_3$} such that \footnote{This relation
requires the consistency condition $d\,(KJ_8)={1\over 2}\,\chi_b\,J_9$, which
is indeed true, but to prove it one can not use naively Leibnitz's rule, see
appendix  $B$.}
\beq
\label{ll11}
KJ_8+{1\over 2}\,\chi_b^0\,J_9=d\,(X J_8).
 \eeq
A proof of this relation, together with an explicit expression for $X$,
are reported in appendix $B$. For our purposes the defining
relation \eref{ll11} is sufficient, because it determines all
properties of $XJ_8$ modulo a closed form.

Another important point regarding $L_{12}$ is that the
Chern--Simons forms $P_7^0$ and $\chi_b^0$ appear evaluated
respectively on the worldvolumes $M_6$ and $M_2$, so that no
unphysical extensions of the $SO(5)$-- and
$SO_b(4)$--connections show up.

Eventually we can also give an explicit expression for the
Wess--Zumino action, written in eleven--dimensional space
($dL_{11}=L_{12}$),
 \bea
\label{l11}
 L_{11}&=&{1\over 6}\,BdBdB +{g\over 2}\,BdBK
 -{g\over 2}\,b\left(dB+{e\over g}\,j_4\right)J_5
+g^2B\left(I_8+{1\over 2}KK\right) \nonumber\\
 &&+e\,B\left(J_8+{1\over 2}\,k\,J_5\right)
+ g^3\left(I_7^0+{1\over 24}(P_7^0+Y_7)\right)K +eg\,XJ_8,
 \eea
where the invariant form $Y_7$ is defined in \eref{y} -- see
\cite{LMT} for an explicit expression --  and we recall that
$KK={1\over 4}d(P_7^0+Y_7)$.
The $b$--dependence of $L_{11}$ is fixed by the
Bianchi identity \eref{3} and by the PST--symmetries or,
equivalently, by the equation of motion \eref{4}. To derive the
$B$--equation of motion \eref{2} from ${1\over
G}\left(S_{kin}+\int_{M_{11}}L_{11}\right)$, the only non trivial
point is the knowledge of the variation of the Born--Infeld action
under a generic variation of $B$; taking advantage from \eref{4}
one obtains \cite{dbipst},
$$
\delta\int_{M_6}d^6\sigma\sqrt{-g}\,{\cal L}(h)= {1\over 2}
\int_{M_6} h\,\delta B= -{1\over 2} \int_{M_{11}} hJ_5\,\delta B.
$$
The integral $\int L_{11}$ is clearly also invariant under the
ordinary gauge transformations
$$
\delta B=d\Lambda, \quad \delta b=d\lambda -\Lambda|_{M_6}.
$$

\subsection{$Q$--invariance and gravitational anomalies}

The terms in $L_{11}$ which are not fixed by the equations of
motion are the ones multiplying $g^3$ and $eg$ in the second line
of \eref{l11}; they are indeed crucial to establish
$Q$--invariance of the Wess--Zumino action and they introduce, in
turn, gravitational anomalies.

The invariance under $Q$--transformations for $B$ of the
Wess--Zumino action in its eleven--dimensional form is less
obvious than in its manifestly invariant twelve--dimensional form;
for $e=0$ this invariance has been explicitly verified in
\cite{LMT}, proving the necessity of the $g^3$--terms. The
necessity of the term $eg XJ_8$ is again a consequence of
this invariance; under
 $$
K\rightarrow K+dQ,\quad B\rightarrow B-gQ,
 $$
concentrating on the terms proportional to $e$ in $L_{11}$, one is left
with the $Q$--variation
 $$
\Delta \left(e BJ_8\right)=-eg QJ_8,
 $$
which should cancel against the variation of $eg XJ_8$.
The transformation property of $X$ can be deduced from
its defining relation \eref{ll11},
 $$
d \left(\Delta X J_8\right)=dQJ_8=d(QJ_8)-QJ_9=d(QJ_8),
 $$
where $QJ_9$ vanishes because $Q$ vanishes on $M_6$, and hence also
on $M_2$. This gives
 \beq\label{XQ}
\Delta \int_{M_{11}} XJ_8 =\int_{M_{11}}QJ_8,
 \eeq
 and
 $$
\Delta\int_{M_{11}}\left(eBJ_8 +eg XJ_8\right)=0.
 $$
This proves that $XJ_8$ is necessary to restore
$Q$--invariance of $S_{wz}$. On the other hand, \eref{ll11}
implies also that under $SO_b(4)$--transformations one has
 $$
 d(\delta XJ_8) =-{1\over 2}d(\chi_b^1\,J_9),
 $$
and therefore
 \beq
 \label{anoml11}
 \delta\int_{M_{11}}(egXJ_8) =-{eg\over 2}\, \int_{M_2}\chi_b^1,
 \eeq
which represents a gravitational anomaly.

Under $Q$--transformations of $b$ (see \eref{anomb})
 $$
k\rightarrow k+dq,\quad b\rightarrow b -{e\over g}\,q,
\quad q|_{M_2}=\chi_a^1,
 $$
the Wess--Zumino action is invariant up to a gravitational
anomaly, as can be seen by direct inspection,
 \beq\label{anoma}
  \delta \int_{M_{11}}L_{11}={e^2\over 2g}\int_{M_2}\chi_a^1.
 \eeq

Eventually one can compute the gravitational anomalies carried by the
Wess--Zumino action. Using \eref{l12} and \eref{anomb} (or \eref{l11},
\eref{anoml11} and \eref{anoma}) one sees that there are anomalies
supported on $M_6$ and $M_2$,
 \beq\label{anom}
\delta \left({1\over G}\,S_{wz}\right) =-{g^3\over
G}\int_{M_6}\left(I_6^1+{1\over 24}\,P_6^1\right)-{eg\over
2G}\int_{M_2}\left(\chi_b^1-{e\over g^2}\,\chi_a^1\right).
 \eeq
These classical anomalies cancel just against the quantum ones,
\eref{m5anom} and \eref{m1anom}, if the charges satisfy
\eref{minimal}.

We may rephrase this result as follows. Once we have introduced
the Chern--kernels $K$ and $k$, it is straightforward to write a ``minimal''
classical action which gives rise to the equations of motion
\eref{2} and \eref{4}, since all terms depending on $B$ and $b$ are fixed.
This action is
${1\over G}\left(S_{kin}+\int {\widetilde L}_{11}\right)$,
where ${\widetilde L}_{11}$ is given in \eref{l11} but without the terms
in $g^3$ and $eg$. This would lead to the total effective action
(classical + quantum)
$$
{\widetilde \Gamma}={1\over G}\left(S_{kin}+\int {\widetilde L}_{11}\right)
 +\Gamma_q.
$$
Since ${1\over G}\int {\widetilde L}_{11}$ carries already the
$SO_a(4)$--anomaly polynomial $\pi \chi_a$ on $M_2$, the effective action
${\widetilde \Gamma}$ is plagued by the gravitational anomalies
$2\pi\left(I_8+{1\over 24}P_8\right)$ on $M_6$ and $\pi \chi_b$ on $M_2$,
but also by $Q$--anomalies.
What we have shown in this paper is the non--trivial fact that there
exists the counterterm
 $$
2\pi\int_{M_{11}}\left(I_7^0+{1\over 24}(P_7^0+Y_7)\right)K
+2\pi \int_{M_3}X,
 $$
which {\it cancels the gravitational  and $Q$--anomalies simultaneously}.

The values \eref{minimal} for the charges,
which ensure anomaly cancellation, satisfy in particular the
Dirac--quantization condition $eg=2\pi nG$ with $n=1$; this is,
actually, a remarkable coincidence since the Dirac--quantization
results in general from an unobservability--requirement of the
Dirac--brane, but in the present case no Dirac--brane at all was
introduced, because the $M2$--brane has a boundary.
The physical meaning of this coincidence can be traced back to the
limiting case in which the open $M2$--brane becomes a closed one
and leaves the $M5$--brane. Since the $M2$--charge remains
unchanged during this process and since a closed $M2$--brane is
consistent if its charge satisfies the Dirac--quantization
condition, one expects that our Wess--Zumino action for an open
membrane reduces to the one of a closed one, if the boundary of
the membrane shrinks to zero. In particular, in this limit a
Dirac--brane for $M2$ should in turn appear. This happens, indeed, as
follows. As the $M2$--brane closes one has to take simultaneously
the limits
$$
j_4\rightarrow 0,\quad J_9=dJ_8\rightarrow 0,\quad k\rightarrow 0.
$$
The field strength $h$ reduces to $db+B|_{M_6}$, and the unique
contribution of order $e$ surviving in \eref{l12} is $eHJ_8$. It
contributes to the classical action with
 $$
S_e={e\over G}\int_{M_{12}}HJ_8.
 $$
Since the membrane is now closed, $dJ_8=0$, we can introduce a
Dirac--brane and write $J_8=dC_e$. Using $eg=2\pi G$ the
interaction above can then be rewritten as
 $$
S_e={e\over G}\int_{M_{11}}HC_e+2\pi\int_{M_{12}}J_5C_e.
 $$
The second term is an (irrelevant) integer multiple of $2\pi$ and
the first term reproduces the standard interaction between closed
branes and dual branes, see \eref{I2}. The integral $\int L_{11}$ reduces
correspondingly to the Wess--Zumino action for interacting closed $M2$--
and $M5$--branes, constructed in \cite{LMT}.

We end this section with a comparison of our approach with the one
of \cite{Harvey}, adapted according to \cite{BK,BHR}, in absence of $M2$.
Specifically we may compare our Wess--Zumino term  \eref{l11} 
-- with $e=0$ and disregarding the obvious terms containing $I_8$
and $I_7^0$ -- with the Chern--Simons term $S_{SC}$, eq. (19) of ref. 
\cite{BK}. In that approach the integration is performed over $M_{11}$
minus a tubular neighborhood of $M_6$ and, restricted to this space, 
$K={1\over 2}\,e_4$ is a closed form, hence locally $e_4=de_3$. In this 
framework, by construction, $S_{CS}$ looses terms supported on $M_6$ 
and one can see that $L_{11}$ differs from $S_{CS}$ indeed by terms 
proportional to $J_5$. These terms are actually needed to get the correct
equation of motion \eref{2} for $B$. 

There remains, however, a difference between $S_{CS}$ and $L_{11}$ also in 
the bulk, regarding the anomaly cancelling term; this is written in \cite{BK}
as ${1\over 8}e_3e_4e_4$, and in $L_{11}$ as ${1\over 4}K_4(P_7^0+Y_7)$.
These two eleven--forms can be mapped into one another through a {\it formal}
integration by parts, using $K_4={1\over 2}de_3$. However, due to the presence
of the tubular neighborhood this operation leads to additional 
terms in the action, and it is not clear whether the two cancellation 
mechanisms are equivalent or not. In our opinion this point requires further
investigation.

\section{$B_3\leftrightarrow B_6$ duality--invariant action}

In this section we present a classical action, equivalent to the
one constructed in the previous section, which involves the
three--form potential $B$ -- now we call it $B_3$ and the field
strength $H_4=dB_3+gK$ -- and its dual six--from potential $B_6$,
in a duality symmetric and manifestly Lorentz--invariant way. For
the problems regarding such a construction when one uses
Dirac--branes instead of Chern--kernels see reference \cite{bbs}.
The knowledge of a consistent duality--invariant action may be
useful for various purposes, for example for the investigation of
the flux quantization conditions of the dual curvature $H_7$, or
for an analysis of dimensional reductions involving dual branes
and dual potentials. Eventually such an action allows
also a comparison with the $M5$--brane $\sigma$--model action
\cite{prl,schwarz} (where the supergravity  fields are treated
as source--less i.e. they satisfy the equations of motion of pure
supergravity), because also the construction of the $\sigma$--model action
involves necessarily $B_3$ as well as $B_6$.

\subsection{PST--action for $b$}

We rewrite the classical action of the previous section
(disregarding the kinetic terms for the metric and for the
$M2$--brane, which are irrelevant for the present purposes)
 \beq\label{action}
 S_0[B_3,b,c]=-{1\over 2}\int_{M_{11}}H_4*H_4 -g\int_{M_6}d^6\sigma\sqrt{-g}
\,{\cal L}(h)+\int_{M_{11}}L_{11}.
 \eeq
In this subsection
we review briefly the basic ingredients of the PST--approach for the
self--interacting chiral two--form $b$ \cite{dbipst}, specifying 
in particular the lagrangian ${\cal L}(h)$. One introduces
an auxiliary scalar field $c(\sigma)$ on $M_6$, and constructs
the unit vector
 $$
 v_i={\pa_i c\over \sqrt{\pa c\,\pa c}},\quad v^iv_i=1.
 $$
Then one can define two  two--forms on $M_6$, contracting $h$ and $*h$ with
this vector:
 $$
{\cal H}\equiv i_vh,\quad \widetilde {\cal H}\equiv i_v*h,
\quad h=-(v{\cal H}+*v\widetilde {\cal H}),
 $$
where $v$ is the one--form $dc/\sqrt{\pa c\,\pa c}$.
In terms of these data one has
 $$
{\cal L}(h)={\cal L}_{BI}(\widetilde {\cal H}) -
{1\over 4}{\cal H}^{ij}\widetilde {\cal H}_{ij},
\quad
{\cal L}_{BI}(\widetilde {\cal H})=
\sqrt{det(\delta_i{}^j+i\widetilde {\cal H}_i{}^j)},
 $$
and the precise form of the equation of motion \eref{4} is
 \beq
\label{precise}
{\cal H}={\cal W}(\widetilde {\cal H})\equiv d\sigma^id\sigma^j\,
{\delta{\cal L}_{BI}\over\delta \widetilde {\cal H}^{ij}}.
 \eeq
Since we have ${\cal W}(\widetilde {\cal H})=\widetilde {\cal H}
+o(\widetilde {\cal H}^3)$, and
${\cal H}=\widetilde {\cal H}$ is the same as $h=*h$, \eref{precise} is
indeed a non--linear version of $h=*h$.
After fixing the PST--symmetries, see below, the action $S_0$
gives rise to \eref{precise}. This equation of motion can indeed
be rewritten in a manifestly Lorentz--invariant way, and the PST--symmetries
ensure correspondingly that $c$ is non propagating.

\subsection{The potential $B_6$ and its curvature $H_7$}

The first step in writing a duality--invariant action consists
in the introduction of a dual potential, i.e. in the solution
of the $B_3$--equation of motion \eref{2} in terms of a potential. Once this
is done in a consistent way, the construction of the duality--invariant
action a la PST is almost canonical.

A consistent reconstruction of a potential $B_6$ relies on the fact
that the r.h.s. of \eref{2} is a well defined {\it invariant closed} form,
so it can be written as the differential of some seven--form. To make this
seven--form explicit we must introduce a new target--space form $O_7$ whose
existence is guaranteed by
 $$
d(kJ_5+J_8)=-j_4J_5+J_9=0\rightarrow kJ_5+J_8=dO_7.
 $$
This allows to recast the Bianchi identity \eref{1} for $B_3$ and its
equation of motion \eref{2}, in the equivalent system of equations
 \bea
H_4&=&dB_3+gK,\label{H4}\\
H_7&=&dB_6+{1\over 2}B_3\,dB_3+g(d\widehat b +B_3)K+
g^2\left(I_7^0+{1\over 8}(P_7^0+Y_7)\right)+e\,O_7,\label{H7}\\
  H_4&=&*H_7.
 \eea
The dual potential is defined a priori up to a field redefinition;
the choice \eref{H7} corresponds to a choice for $B_6$ which is {\it regular
on $M_6$}, as we will see now. The necessity of a regular $B_6$ results from
the fact that the duality--invariant action will contain the minimal coupling
$\int B_6J_5=\int_{M_6}B_6$.

In the formula for $H_7$ we introduced a field $\widehat b$ which represents
an unphysical (holographic) target--space extension of the two--form 
potential $b$ on $M_6$,
 $$
  \widehat b|_{M_6}=b,
 $$
so let us first establish that the extension is unobservable. Changing the
extension we have
 $$
 \widehat b^\prime =\widehat b + \Delta\widehat b,
\quad \Delta\widehat b|_{M_6}=0,
 $$
which leads in $H_7$ to a change $gd(\Delta\widehat b) K
=gd(\Delta \widehat b K)$.
$H_7$ is kept invariant if $B_6$ changes accordingly by
 \beq
 \label{suppl}
 B_6^\prime-B_6 =-g\Delta\widehat b\,K.
 \eeq
Notice that this transformation keeps $B_6|_{M_6}$ invariant
\footnote{There exists an alternative definition of a six--form potential
which avoids the extension of $b$: you can choose the singular potential
$\widetilde B_6= B_6+g\widehat b K$. This gives
 $$
H_7=d\widetilde B_6+{1\over 2}B_3dB_3-gbJ_5+gB_3K+
g^2\left(I_7^0+{1\over 8}(P_7^0+Y_7)\right)+eO_7,
 $$
and no unphysical extension of $b$ shows up. But with this potential
the formula for $H_7$ shows up a $\delta$--like singularity
on $M_6$ (the term $bJ_5$),
which is canceled by the singularity contained in $\widetilde B_6$.
Use of the potential $B_6$ and of an extended $\widehat b$ makes this
cancellation explicit.}.

The field--strength $H_7$ contains necessarily singularities
near $M_6$ and, as in the case of the four--form field--strength $H_4$, these
singularities have to be universal \footnote{The pullback of $H_7$ to
$M_6$ vanishes for dimensional reasons, so the singularities we are speaking
about are considered in the algebraic sense.}.
To make them explicit we should know the singular behaviour of $O_7$ near
the $M5$--brane. From
its defining relation -- which determines it modulo a closed form -- it is
clear that there exists a choice, such that its singular part near $M_6$ is
$kK$. Since $B_6$, $B_3$, $I_7^0$ and $P_7^0$ are regular near $M_6$,
the universal singular behaviours of the curvatures are given by
 \bea
\left[H_4\right]_{sing} &=& gK, \nonumber\\
\left[H_7\right]_{sing} &=& g\,hK+{g^2\over 8}\,Y_7.\nonumber
 \eea

The consistency of this construction requires now that under
$Q$--transformations of $B_3$, $H_7$ is invariant in compatibility with a
{\it regular transformation law} for $B_6$ on $M_6$ \footnote{Since the
r.h.s. of
\eref{2} is $Q$--invariant, it is obvious that there exists always a
transformation $\Delta B_6$ which keeps $H_7$ invariant; it is less trivial
that $\Delta B_6$ is finite on $M_6$.}. Due to the presence of the
$SO(5)$--Chern--Simons form $P_7^0$ in $H_7$, one expects that under
$Q$--transformations one has the remnant anomalous $SO(5)$--transformation
 \beq\label{b6trans1}
 \Delta B_6|_{M_6}=-{g^2\over 8}P^1_6|_{M_6}.
 \eeq

Since $O_7$ is $Q$--invariant, to deduce the transformation of $B_6$
one must first determine the transformation law of
 $$
S\equiv P_7^0+Y_7.
 $$
Under $K^\prime =K+dQ$, since $KK={1\over 4}\,dS$, one obtains easily
 \beq
 \label{Strans}
S^\prime=S+8KQ+4Q\,dQ+d\,{\cal P},
 \eeq
for some six--form ${\cal P}$. But we have also
$S^\prime=P_7^{0\prime}+Y_7^\prime$, and since $Y_7^\prime$ and
$Y_7$ have the same (singular) behaviour near $M_6$ we get
$(Y_7^\prime -Y_7)|_{M_6}=0$, and therefore
 $$
 (S^\prime-S)|_{M_6}=(P_7^{0\prime}-P_7^0)|_{M_6}=dP_6^1|_{M_6}.
 $$
Comparing this with \eref{Strans} and remembering that $Q|_{M_6}=0$ we get, 
apart from a closed form,
 \beq
 \label{qpull}
{\cal P}|_{M_6}=P_6^1|_{M_6}.
 \eeq
Using \eref{Strans} it is  eventually straightforward to show that $H_7$ is
$Q$--invariant under
 \bea
   K^\prime&=&K+dQ,\\
   B_3^\prime&=&B_3-gQ,\\
   B_6^\prime&=&B_6-g\left(d\widehat b+{1\over 2}B_3\right)Q
       -{g^2\over 8}\,{\cal P},
 \eea
which proves also that $B_6$ has a regular transformation law on
$M_6$, given by \eref{b6trans1}. Under $SO(1,10)$--transformations
one must obviously also have
 \beq\label{b6trans2}
\delta B_6=-g^2 I_6^1.
 \eeq
$H_7$ is also invariant under ordinary gauge transformations,
 \beq
\label{ordinary} \delta B_3=d\Lambda_2,\quad \delta
B_6=d\Lambda_5-{1\over 2}\,\Lambda_2\,dB_3,\quad \delta\widehat
b=d\Lambda_1-\Lambda_2.
 \eeq

\subsection{Duality--invariant action and PST--symmetries}

Once we have introduced a dual potential the construction of a
duality--invariant action, according to the PST--approach, requires the
introduction of an auxiliary target--space scalar field $C(x)$. Due to
the presence of a chiral field on the $M5$--brane, whose action required
already an auxiliary field $c(\sigma)$ on $M_6$, as shown in \cite{bbs}
the consistency of the whole construction demands now that $c$ is the
pullback of $C$,
 $$
 c=C|_{M_6}.
 $$
Apart from this, the construction proceeds as follows. Define the
target--space vector
 $$
V_\mu={\pa_\mu C\over \sqrt{\pa C\,\pa C}},\quad V^\mu V_\mu=1,
 $$
and the target--space three-- and six--forms
 $$
f_3\equiv i_V(H_4-*H_7),\quad f_6\equiv i_V(H_7-*H_4),
 $$
which realize the decomposition
 \beq\label{decomp}
   H_4-*H_7=-(Vf_3+*Vf_6), \quad V=dC/\sqrt{\pa C\,\pa C}.
 \eeq
Then the duality--invariant action is given in terms of \eref{action} by
 \beq\label{duinv}
S[B_3,B_6,\widehat b,C]=S_0[B_3,b,c]
+{1\over 2}\int_{M_{11}}f_3*f_3.
 \eeq
In this new framework $H_4$ and $H_7$ are defined in terms of $B_3$ and $B_6$
as in \eref{H4} and \eref{H7}, and the duality relation $H_4=*H_7$ is
promoted to an equation of motion which should be produced by $S$ together
with the $b$--equation \eref{precise} 
\footnote {Despite the explicit appearance of $\widehat b$ in $H_7$ the
action $S$ is actually only a functional of $b$, as explained in the previous
subsection.}.

We show now that $S$ entails the same dynamics as $S_0$, i.e. that the two
actions are physically equivalent. The starting point are
the equations of motion for $B_3$, $B_6$, $b$ and $C$; the complete set,
and their derivation, is given in appendix C. For our purposes it is
sufficient to know the equations for $B_6$ and $B_3$, which read
 \bea
 d(Vf_3)&=&0, \label{b6}\\
d(Vf_6)&=&gv({\cal H}-{\cal W})J_5 -Vf_3H_4,\label{b3}
 \eea
where ${\cal H}$ and ${\cal W}$ are defined in subsection 6.1.
Knowledge of these equations is sufficient because the $b$--equation is
implied by the $B_3$--equation (as
a direct consequence of the gauge invariance $\delta B_3=d\Lambda_2$,
$\delta b=-\Lambda_2|_{M_6}$), and the
the $C$--equation is identically satisfied if the other three equations of
motion hold. This is a consequence of the fact that $C$ is a non--propagating
auxiliary field. The action $S$ is in fact invariant 
(see appendix C for the proof) under the
PST--transformations $(I)$ 
 \beq\label{uno}
\delta \widehat b=dC\Psi_1,\quad\delta B_3=dC \Psi_2,
\quad \delta B_6=dC\Psi_5,\quad\delta C=0,
 \eeq
with transformation parameters $\Psi_1$, $\Psi_2$ and $\Psi_5$,
and under $(II)$
 \bea
\delta C&=&\Phi,\\
\delta \widehat b&=&
\Phi\left[{1\over \sqrt{\pa c\pa c}}\,({\cal H}-{\cal W})\right],\\
\delta B_3&=&{\Phi\over\sqrt{\pa C\pa C}}\,f_3,\\
\delta B_6&=&{\Phi\over\sqrt{\pa C\pa C}}\,f_6-{1\over
2}B_3\,\delta B_3 -g\,\delta \widehat b\,K,\label{due}
 \eea
where $\Phi$ is a scalar transformation parameter; with
$\left[{1\over \sqrt{\pa c\pa c}}({\cal H}-{\cal W})\right]$ we indicate an
arbitrary target--space extension of the two--form between brackets.

$\Phi$ being arbitrary, the symmetry $(II)$ ensures that the auxiliary field
$C$ is non propagating, and the symmetries $(I)$
allow to reduce the equations of motion \eref{b6} and \eref{b3} to
 $$
H_4=*H_7,\quad  {\cal H}-{\cal W}=0,
 $$
which proves that $S$ and $S_0$ are equivalent. To see how this happens
consider the most general solution of \eref{b6},
 $$
 Vf_3=dCd\widetilde \Psi_2,
 $$
for some two--form $\widetilde \Psi_2$. Since under a transformation
$(I)$ of $B_3$ we have $Vf_3\rightarrow Vf_3-dCd\Psi_2$, we can use this
symmetry (with $\Psi_2=\widetilde\Psi_2$) to set $Vf_3=0$. This is the same
as
 $$
f_3=0,
 $$
because $i_Vf_3=0$ identically. At this point, taking the differential
of \eref{b3} one gets \footnote{The $b$--equation of motion is precisely
obtained by taking the differential of the $B_3$--equation, see appendix C.}
 $$
d(v({\cal H}-{\cal W}))=0,
 $$
and one can use the same procedure as above --
using the symmetry $(I)$ of $\widehat b$ -- to reduce this equation to
 $$
{\cal H}-{\cal W}=0.
 $$
This simplifies eventually the $B_3$--equation to $d(Vf_6)=0$,
and the symmetry $(I)$ of $B_6$ can be used to reduce it further to $f_6=0$.
The identity \eref{decomp} completes then the proof.

\subsection{Comparison with the $M5$--brane $\sigma$--model}

The action presented in the present paper describes a fully interacting theory
of dynamical supergravity with dynamical brane--like sources, as is appropriate
to discuss, for example, anomaly cancellation. On the contrary
the $M5$--brane $\sigma$--model action \cite{prl,schwarz}
constrains the supergravity fields to satisfy the (source--less)
equations of motion of pure $D=11$ Sugra, as required by
$\kappa$--symmetry. In the comparison this fundamental difference has to be
taken into account.

The bosonic part of the $\sigma$--model action reads
 $$
   S^\sigma =-g\int_{M_6}d^6\sigma\sqrt{-g}\,{\cal L}(h)+
 g\int_{M_6}\left(B_6-{1\over2}\,b\,dB_3\right),
  $$
where $h=db+B_3|_{M_6}$, since we disregard here the $M2$--brane
($e=0$). The structure of the Wess--Zumino term (the second integral) 
is fixed by invariance
under the ordinary gauge transformations \eref{ordinary}. Since
the $\sigma$--model action contains as main building block the standard 
minimal coupling term $\int_{M_6} B_6=\int_{M_{11}}J_5B_6$, to make the
comparison we should view our action $S$ from the
$B_6$--point--of--view. This can be done using the identity
 $$
 -{1\over 2}\left(H_4*H_4-f_3*f_3\right)=
{1\over 2}\left(H_7*H_7+f_6*f_6\right)-H_4H_7,
 $$
which is once more a consequence of \eref{decomp}. This allows to rewrite $S$
as
 \bea
S&=&\widehat S_{kin}+\widehat S_{wz},\nonumber\\
\widehat S_{kin}&=&{1\over
2}\int_{M_{11}}\left(H_7*H_7+f_6*f_6\right)
-g\int_{M_6}d^6\sigma\sqrt{-g}\,{\cal L}(h),\nonumber\\
\widehat
S_{wz}&=&\int_{M_{11}}\left(L_{11}-H_4H_7\right)\nonumber.
 \eea
This form of the action privileges indeed the role of the
potential $B_6$: it appears only through the canonical kinetic
term ${1\over 2}H_7*H_7$ (apart from the square of the six--form
$f_6$ which vanishes on--shell), and through the standard minimal
coupling to the $M5$--brane contained in $H_4H_7$. The kinetic
term for the $M5$--brane in $\widehat S_{kin}$ is already the same
as in $S^\sigma$, and the Wess--Zumino action $\widehat S_{wz}$ can 
be split in the three contributions,
 \bea
\widehat S_{wz}&=&S_{wz}^{sugra}+S_{wz}^\sigma+S_{wz}^{int},\nonumber\\
         S_{wz}^{sugra}&=&-{1\over 3}\int_{M_{11}}B_3dB_3dB_3,\nonumber\\
         S_{wz}^\sigma&=&g\int_{M_6}\left(B_6-{1\over2}\,b\,dB_3\right),
\nonumber\\
         S_{wz}^{int}&=&-g\int_{M_{11}}\left(B_3dB_3K+gB_3KK
             +{g^2\over
             12}\,(P_7^0+Y_7)K\right)+g\int_{M_6}b\,dB_3,\nonumber
 \eea
where all three contributions are separately invariant under the
ordinary gauge transformations \eref{ordinary}. $S_{wz}^{sugra}$
represents the pure supergravity contribution: the coefficient
$-1/3$ instead of $+1/6$ of the standard $B_3$--based Sugra is due
to the fact that this scheme privileges $B_6$. $S_{wz}^\sigma$ is
indeed the $\sigma$--model Wess--Zumino action showing up in
$S^\sigma$.  Very naively one could have expected that the total
Wess--Zumino action is simply $S_{wz}^{sugra}+S_{wz}^\sigma$, but
the fact is that  while $S_{wz}^\sigma$ is $Q$--invariant (modulo
gravitational anomalies) $S_{wz}^{sugra}$ is not. The additional
interaction terms of $S_{wz}^{int}$  (of order $g$, $g^2$, and
$g^3$) cure this non--invariance. By direct inspection one sees
indeed that
 $$
S_{wz}^{sugra}+S_{wz}^{int}=-2\int_{M_{11}}
\left(L_{11}-g^2H_4\,I_7^0\right),
 $$
which is manifestly $Q$--invariant (modulo gravitational
anomalies). Eventually, it is only the total action, Wess--Zumino
+ kinetic terms, that is invariant under PST--symmetries.

Finally it is instructive to  analyze how the duality--invariant
action realizes the cancellation of the gravitational $M5$--brane
anomalies. $S_{wz}^{sugra}$ is invariant and $S_{wz}^\sigma$
carries the anomaly (see \eref{b6trans1} and \eref{b6trans2}),
 $$
 -\left(I_8+{1\over 8}\,P_8\right),
 $$
but this would not be the right amount to cancel the quantum
anomaly $I_8+{1\over 24}\,P_8$. But also $S_{wz}^{int}$ carries an
anomaly, ${1\over 12}\,P_8$, and it ensures hence the
matching; gravitational anomalies and $Q$--invariance are once
more entangled.

Our duality invariant--action $S$ can be compared also with the 
duality--invariant action $S_{BBS}$ of \cite{bbs}, formula (5.28), 
that is based on 
a Dirac--brane approach. In that action the (ill--defined) products of
$\delta$--functions 
on  Dirac--branes -- like $C_gC_g$ -- are ignored. This implies that $S_{BBS}$ 
contains at most terms linear in $g$ and that it carries no gravitational 
anomalies. Correspondingly one can see that our action \eref{duinv} 
(for $e=0$) reduces to $S_{BBS}$ -- after correcting a factor 1/2 -- if 
one neglects all terms of order $g^2$ 
and $g^3$, and replaces the Chern--kernel $K$ with the $\delta$--function
on the Dirac--brane $C_g$.

\section{Concluding remarks and open problems}

There are attempts in the literature to construct low energy effective
actions for $M5$--branes (with or without $M2$--branes) 
relying on potentials $B_3$ which are 
ill--defined on the $M5$--brane worldvolume \cite{bbs,alwis,jussi,xavier};
they are based on Dirac--branes, or something equivalent to them, and 
a consistent treatment of the resulting singularities is still missing 
\footnote{Since in the $M5$--brane effective action the interaction is cubic 
and since products of $\delta$--functions do not define distributions 
the authors think that for this case a consistent
treatment can not be achieved.}. 
These approaches refrain, in particular, from specifying the singular 
behaviour of the field strength $H_4$ near the $M5$--brane and they entail 
correspondingly two main drawbacks:
the dynamics i.e. the action -- but even the equations of motion -- contain 
ill--defined objects, such as the pullback of $B_3$ or -- even worse -- 
of $H_4$ and, moreover, the
(formal) actions which result from these approaches do not cancel the
gravitational anomalies. The Chern--kernel approach solves these
two problems simultaneously.   

In this paper we considered a topologically trivial space--time. 
Nevertheless the Wess--Zumino action
$$ 
S_{wz}=\int_{M_{12}}L_{12},
$$
with $L_{12}$ given in \eref{l12}, should make sense also in a target--space
$M_{11}$ with non--trivial topological properties. In such a space--time the 
crucial feature is the dependence of $S_{wz}$ on the chosen twelve--manifold
$M_{12}$ with boundary $M_{11}$. In the absence of {\it global} quantum 
anomalies one would require that ${1\over G}S_{wz}$ is independent of the 
chosen 
$M_{12}$ modulo $2\pi$. This would be equivalent to require that
$$
{1\over G}\int_M L_{12}=2\pi n,
$$ 
with $n$ integer, for a generic closed manifold $M$. However, as shown
in \cite{flux1,flux2}, even  in the absence of $M2$-- and $M5$--branes the 
above integral is in general only an integer multiple of $\pi$, meaning that
$exp\left({i\over G}S_{wz}\right)$ carries a residual sign--dependence on 
$M_{12}$. But in \cite{flux1} it is 
then also shown that the quantum effective action carries a global anomaly,
due to the fermionic determinant of the Rarita--Schwinger operator, 
which compensates precisely this sign--ambiguity. 

A similar situation is supposed to arise in the presence of $M2$--
and $M5$--branes. In this case there are additional global quantum
anomalies \cite{Witten,flux2} due to the branes, but one expects again that
the total effective action is well--defined. This conjecture needs still
to be proved, but since the formula \eref{l12} is explicit and involves 
only objects with a simple and clear geometrical meaning, we hope to be able
to provide a proof based on that formula.

In the absence of branes the proof of cancellation of global anomalies 
of reference \cite{flux1} relied on the statement that $H$ 
defines a half--integral cohomology class, i.e. $[H]\in {g\over 2}\,{\bf Z}$.
In presence of an $M5$--brane a necessary condition for global anomaly 
cancellation is then that $H$ represents a half--integral
class {\it in the complement of $M5$}. We can check here that 
this minimal condition can be easily satisfied in our construction.
The first step concerns the definition of $H$ in terms of the
Chern--kernel $K$. In a topologically non--trivial space--time the 
form $K$ must have the following defining properties: 1) near $M5$ it 
behaves as in \eref{even} and 2) in the complement of $M5$ it amounts
to a closed form. An explicit realization like \eref{even} is 
still available  
if there exists a map $y^a(x)$ from  target--space to ${\bf R}^5$, reducing to
normal coordinates in a tubular neighborhood of $M5$ and nowhere vanishing
in its complement. Defining 
$\yh^a=y^a/|y|$ we can then introduce a globally defined four--form $K$ 
according to \eref{even} and write
 \beq\label{coho}
H=H_0+gK,
 \eeq
where $H_0$ (locally $dB_3$) is a half--integral cohomology class,  
$[H_0]\in {g\over 2}\,{\bf Z}$, as assumed in \cite{flux1} in absence of
branes. Using the above
realization for $K$ we can now show that also
$$ 
[K]\in {\bf Z}/2\quad {\rm in\,\, the\,\, complement\,\, of}\,\, M5.
$$
Take in each point of target--space an $SO(4)$--basis $n^{ar}(x)$ 
normal to $\yh^a$, $n^{ar}\yh^a=0$, $n^{ar}n^{as}=\delta^{rs}$, 
$r,s=(1,\cdots,4)$.
Construct then the $SO(4)$--connection and curvature 
$W^{rs}\equiv n^{sa}Dn^{ra}$, 
$T=dW+WW$. Since one has 
$T^{rs}=n^{ra}n^{sb}\left(F^{ab}+D\yh^aD\yh^b\right)$
and $\ve^{a_1\cdots a_5}\yh^{a_5}=\ve^{r_1\cdots r_4}n^{r_1a_1}\cdots
n^{r_4a_4}$, it is immediately seen that
$$
K={1\over 2}\,\chi(T),
$$
i.e. one--half of the $SO(4)$--Euler--form of the curvature $T$
\footnote{The frames $n^{ar}$ can not be glued together continuously all over 
the target--space. Even in a space--time with trivial topology they carry 
necessarily a singularity along a seven--manifold $M_7$ whose boundary
 is $M5$. This 
implies that $T$ is singular along $M_7$ while $\chi(T)$, being 
$SO(4)$--invariant i.e. independent of $n^{ar}$, is singular only
on $M5$. This means that we have $d\chi(T)=0$ in the complement of $M5$,
while in the whole target--space we have $d\chi(T)=2J_5$.}.
This implies that $K$ has in general semi--integer integrals over closed 
four--manifolds which do not intersect $M5$. However, its integral over a 
four--sphere linking $M5$ equals unity, since the Euler--characteristics of a 
four--sphere is two: this reflects clearly the fact that the $M5$--brane
carries one unit of magnetic charge. 

\section{Appendices}

\appendix

\section{Regularizations}

Instead of giving general formulae for the ``real algebraic approximation
mode'' \cite{charact}, we present here examples one for an even
and one for an odd current, generalizations being straightforward.

First we approximate our $\delta$--like currents $J$ with smooth ones
$J^\ve$ that preserve $SO(n)$--invariance and depend algebraically
on $F^{ab}$, $y^a$ and $Dy^a$; for $n=5$ and $n=4$  we have
 \bea
J_5^\ve&=&{1\over 4(2\pi)^3}\,{\ve\over \ve^2+y^2}\,\ve^{a_1\cdots a_5}
           Dy^{a_1}\left(F^{a_2a_3}F^{a_4a_5}+
          {4\over 3 (\ve^2+y^2)}\,F^{a_2a_3}Dy^{a_4}Dy^{a_5}\right.
\nonumber\\
&&\phantom{{1\over 4(2\pi)^3}{\ve\over \ve^2+y^2}\,\ve^{a_1\cdots a_5}
           Dy^{a_1}}
\left.+{8\over 15(\ve^2+y^2)^2}\,Dy^{a_2}Dy^{a_3}Dy^{a_4}Dy^{a_5}
          \right),
\label{j5reg}\\
J_4^\ve&=&{1\over8(2\pi)^2}{\ve\over
           \sqrt{\ve^2+y^2}}\,\ve^{a_1\cdots a_4}\left(F^{a_1a_2}F^{a_3a_4}+
           {2\over\ve^2+y^2}\,F^{a_1a_2}
           Dy^{a_3}Dy^{a_4}\right.\nonumber\\
           &&\phantom{{1\over8(2\pi)^2}{\ve\over \sqrt{\ve^2+y^2}}\,
           \ve^{a_1\cdots a_4}}
\left.+{1\over (\ve^2+y^2)^2}\, Dy^{a_1}Dy^{a_2}Dy^{a_3}Dy^{a_4}\right).
\label{j4reg}
 \eea
These currents satisfy
 \bea\nonumber
\lim_{\ve\rightarrow 0}J^\ve&=& J\\
dJ^\ve&=&0\nonumber\\
\int_{M_n}J^\ve&=&1,\nonumber
 \eea
where the last integral is along an $n$--dimensional manifold at
$\sigma^i=\sigma_0^i$. The currents $J^\ve$ enjoy therefore the following
properties: they are closed and
invariant, they entail the same total charge as $J$ and they are regular at
$M$: we have indeed the pullback formulae
 \bea\label{a1}
J^\ve_4|_M&=&\chi_4(f)\\
J^\ve_5|_M&=&0,\label{a2}
 \eea
where $\chi(f)$ for a generic even $n$ is the $SO(n)$--Euler--form
on $M$. These formulae represent, in particular, an explicit realization of
the Thom--isomorphism \cite{BT}. Regularized Chern--kernels satisfy by
definition
$$
J^\ve=dK^\ve, \quad \lim_{\ve\rightarrow 0}K^\ve=K,
$$
and it is not difficult to find explicit expressions. For the even kernel
one gets
 \beq\label{k4reg}
K_4^\ve={1\over 4(2\pi)^3}\,\ve^{a_1\cdots a_5}
           \yh^{a_1}\left(f_1F^{a_2a_3}F^{a_4a_5}+
         f_2F^{a_2a_3}D\yh^{a_4}D\yh^{a_5}
            +f_3D\yh^{a_2}D\yh^{a_3}D\yh^{a_4}D\yh^{a_5}\right),
 \eeq
with
 \bea\nonumber
f_1&=&\arctan{y\over\ve}\\
f_2&=&2\left(\arctan{y\over\ve}-{\ve y\over\ve^2+y^2}\right)\nonumber\\
f_3&=& \arctan{y\over\ve}-{\ve y\over\ve^2+y^2}
          -{2\over 3}{\ve y^3\over (\ve^2+y^2)^2},
\eea
and it is straightforward to see that for $\ve\rightarrow 0$ one obtains
\eref{even} for $n=5$. For the odd kernel we get instead
 \beq
K_3^\ve=\Omega_3^\ve+\chi_3^0(A),
 \eeq
where
 \beq
\Omega_3^\ve={1\over 2(4\pi)^2}\,\ve^{a_1\dots
a_4}\yh^{a_1}D\yh^{a_2}\left( g_1 F^{a_3a_4}+
g_2D\yh^{a_3}D\yh^{a_4}\right),
 \eeq
with
 \bea
g_1&=&4\left({\ve\over\sqrt{y^2+\ve^2}}-1\right)\\
g_2&=&{8\over 3}\left({\ve^3+{3\over 2}\ve y^2\over(y^2+\ve^2)^{3/2}}
-1\right).
 \eea
For $\ve\rightarrow 0$ $\Omega_3^\ve$ reduces to \eref{omega}.
The regularized even kernel is invariant, as $K_4$, and the odd one maintains
the transformation property of $K_3$ since $\Omega_3^\ve$ is invariant. Both
kernels are regular on $M$ and one gets
 \bea
K_3^\ve|_M&=&\chi^0_3(a)\\
K_4^\ve|_M&=&0,\label{k4epull}
 \eea
obviously in agreement with \eref{a1}, \eref{a2}.

To illustrate the use of these formulae we provide a simple alternative proof
of \eref{serve}. Considering the limits in the sense of distributions
we can write
$$
d(YK)=\lim_{\ve\rightarrow 0}d\left(YK^\ve\right)=\lim_{\ve\rightarrow 0}
\left(dYK^\ve+YJ^\ve\right),
$$
where we are allowed to use Leibnitz's rule since $K^\ve$ is a regular
distribution. The
first term converges trivially to $dYK$, while the second term is in the limit
necessarily supported on $M$ (as $\ve \rightarrow 0$ only terms with
$y=0$ contribute, see \eref{j5reg}).
This means that
$$
\lim_{\ve\rightarrow 0}YJ^\ve= {\cal Y}J,
$$
for some form ${\cal Y}$ which is defined on $M$ \footnote{Notice that the
existence of the limit $\lim_{\ve\rightarrow 0}YK^\ve=YK$ in the
distributional sense ensures the existence of the limit
$\lim_{\ve\rightarrow 0}d(YK^\ve)$ in the distributional sense; the same
property does in general not hold in linear spaces of functions, endowed with
stronger topology.}. Moreover, since $Y$ as well as $J^\ve$ are
$SO(n)$--invariant forms, also ${\cal Y}$ must be an invariant form on $M$;
but since ${\cal Y}$ is an odd form it must necessarily vanish, because
there are no invariant odd forms on $M$, made out of the $SO(n)$--connection
$a^{ab}$.

Similarly one can show that for even kernels the product $KK$ is a closed form,
$$
d(KK)=\lim_{\ve\rightarrow 0}d\left(KK^\ve\right)=\lim_{\ve\rightarrow 0}
\left(JK^\ve+KJ^\ve\right)=\lim_{\ve\rightarrow 0}KJ^\ve,
$$
where the last equality follows from \eref{k4epull}. As above one has
$\lim_{\ve\rightarrow 0}KJ^\ve={\cal K}J$, for some invariant even form
${\cal K}$ which must be also {\it odd under parity of the normal bundle}
i.e. proportional to $\ve^{a_1\cdots a_n}$, as is $K$. But since $n$ is odd
no such form exists, and hence ${\cal K}=0$.

\section{Existence and construction of the three--form $X$}

We begin giving the explicit expression of the $SO(5)$--kernel
 \beq
\label{k4}
K={1\over 4(4\pi)^2}\,\ve^{a_1\cdots a_5}\,\hat
y^{a_1}\left(F^{a_2a_3}+D\yh^{a_2}D\yh^{a_3}\right)
\left(F^{a_4a_5}+D\yh^{a_4}D\yh^{a_5}\right),
 \eeq
and recalling the definition of the Euler--form, the
associated Chern--Simons form  and its descent, for a generic $SO(4)$--bundle
with connection $W$ and curvature $T=dW+WW$,
 \bea
\chi(T) &=&{1\over 2(4\pi)^2}\,\ve^{r_1\cdots r_4}
T^{r_1r_2}T^{r_3r_4},\nonumber\\
\chi^0(W)&=&{1\over 2(4\pi )^2}\,\ve^{r_1\dots r_4}\left(
W^{r_1r_2}dW^{r_3r_4}+ {2\over 3}\,W^{r_1r_2}\left(
WW\right)^{r_3r_4} \right),\nonumber\\
\chi^1(W)&=& {1\over 2(4\pi )^2}\,\ve^{r_1\dots r_4}
L^{r_1r_2}dW^{r_3r_4},\nonumber\\
 \chi(T)&=&d\,\chi^0(W), \quad \delta \chi^0(W)=d\,\chi^1(W),
  \quad \delta W^{rs}=DL^{rs}.
\nonumber
 \eea
We are searching for a form $X$ on $M_3$ such that
 \beq\label{sempre}
KJ_8+{1\over 2}\,\chi_b^0J_9=d(XJ_8),
 \eeq
where $\chi_b^0(w)$ is the Chern--Simons form of the Euler--form,
with $SO_b(4)$--connection $w^{rs}$ on $M_2$. Since
$KJ_8=K|_{M_3}J_8$, what appears really in this formula is the
pullback of $K$ on $M_3$, and we will now show that the pullback
form is a closed form, more precisely that
 $$
K|_{M_3}=dX,
 $$
where $X$ is the form we are searching for. In $K|_{M_3}$ all
objects of \eref{k4} are evaluated on $M_3$. If we parametrize
$M_3$ by $x^\mu(\sigma)$ we get in particular $u^a(\sigma)\equiv
\yh^a(x(\sigma))$, which determines in each point of $M_3$ a
particular direction. This allows to define on $M_3$ the reduction
-- along $u^a$ -- of the $SO(5)$--connection $A^{ab}$ down to an
$SO(4)$--connection $W^{rs}(\sigma)$. The reduction procedure is
canonical: define on $M_3$ an $SO(4)$--basis $n^{ra}$ normal to
$u^a$, $r=(1,\cdots,4)$,
 $$
n^{ra}u^a=0,\quad n^{ra}n^{sa}=\delta^{rs},\quad u^au^a=1;
 $$
then the reduced connection and curvature are given by
 $$
W^{rs}=n^{sa}Dn^{ra}, \quad T^{rs}=dW^{rs}+W^{rt}W^{ts}.
 $$
With these definitions it is straightforward to show that one has
the identities
 $$
T^{rs}=n^{ra}n^{sb}\left(F^{ab}|_{M_3}+Du^aDu^b\right),\quad
 \ve^{a_1\cdots a_5}u^{a_5}=\ve^{r_1\cdots r_4}n^{r_1a_1}\cdots
  n^{r_4a_4}.
 $$
Using them in the restriction of \eref{k4} to $M_3$, one sees
immediately that this restriction amounts to the Euler--form of
$T$, apart from a factor 1/2,
 $$
 K|_{M_3}={1\over 2}\,\chi(T).
 $$
The three--form $X$ is then simply the Chern--Simons form of $W$,
 \beq
 X={1\over 2}\,\chi^0(W).
 \eeq
To prove that this $X$ satisfies indeed \eref{sempre} it remains
to show that the restriction of $W^{rs}$ from $M_3$ to $M_2$
coincides with the $SO_b(4)$--connection $w^{rs}$. This can be
seen as follows.  Since the $y^a$ are normal coordinates on $M_6$
and $u^a=\yh^a|_{M_3}$, we have that the vector $\widetilde
u^a\equiv u^a|_{M_2}$ coincides with the (normalized) component
normal to $M_6$, of the unique vector on $M_2$ which is tangent to
$M_3$ and normal to $M_2$. But by definition (see section 5.1)
$SO_b(4)$ is the subgroup of $SO(5)$ on $M_2$ which leaves
$\widetilde u^a$ invariant. This means, by construction, that
$W^{rs}|_{M_2}=w^{rs}$.

From the explicit construction of $X$ one can deduce its
transformation properties under $Q$--transformations and
$SO_b(4)$--transformations respectively, checking thus \eref{XQ} and
\eref{anoml11}:
 \bea
 \Delta X&=& Q|_{M_3}+d\phi,\nonumber\\
 \delta X&=& {1\over 2}\,d\,\chi^1(W),
\nonumber
 \eea
where $\phi$ is some two--form on $M_3$ with  $\phi|_{M_2}=0$,
and clearly  $\chi^1(W)|_{M_2}=\chi^1_b$.

\section{PST--invariance}

The main work one has to accomplish is to compute the variation of
$S$ under generic variations of the fields $B_3$, $B_6$, $\widehat
b$ and $C$. The result can be cast in the form
 \bea\nonumber
\delta S &=&\int_{M_{11}}\Bigg[\Bigg(d(Vf_6)+Vf_3H_4
 -gv({\cal H}-{\cal W})J_5\Bigg)\left(\delta B_3+d\delta \widehat b\right)\\
 &&-d(Vf_3)
 \left(\delta B_6+{1\over 2}\,B_3\delta B_3+g K\delta\widehat
 b\right)\nonumber\\
&&+dCd\left({g\over 2\pa c \pa c}\,
 ({\cal H}-{\cal W})({\cal H}-{\cal W})J_5
 -{1\over\pa C\pa C}f_3f_6\right)\delta C\Bigg].\nonumber
 \eea
To arrive at this formula one has to use \eref{decomp} and some
standard relations like $i_V*=*V$ and $*\,i_V=V*$; but, most
importantly, one has to use the identity for four--forms
 $$
 {\cal H}{\cal H}={\cal W}{\cal W},
 $$
which holds  because ${\cal W}$ is the functional derivative of
the Born--Infeld lagrangian. It is indeed the qualifying property
for all allowed lagrangians, i.e. those which give rise to a
consistent (Lorentz--invariant) dynamics for a (self--interacting)
chiral boson in six dimensions.

The equations of motion for $B_6$, $B_3$, $\widehat b$ and $C$ can
be read directly from $\delta S$ and are given respectively by,
 \bea
 d(Vf_3)&=&0\label{A}\\
d(Vf_6)&=&gv({\cal H}-{\cal W})J_5 -Vf_3H_4\label{C}\\
gd(v({\cal H}-{\cal W}))J_5&=&-d(Vf_3H_4)\label{B}\\
dCd\left({g\over 2\pa c \pa c}\,
 ({\cal H}-{\cal W})({\cal H}-{\cal W})J_5
 -{1\over\pa C\pa C}f_3f_6\right)&=&0\label{D}.
 \eea
One sees that the $\widehat b$--equation is supported on $M_6$,
because the r.h.s. of \eref{B} is proportional to $J_5$, thanks to
\eref{A}. This is consistent with the fact that the extension
$\widehat b$ is a pure gauge degree of freedom. It is also
immediate to see that the $\widehat b $--equation is a consequence
of the $B_3$--equation, and that the $C$--equation is a
consequence of the other three equations.

Eventually, proving invariance of $S$ under the
PST--transformations \eref{uno}--\eref{due} -- given the above
form of $\delta S$ -- is a simple exercise.

\bigskip

\paragraph{Acknowledgements.}
The authors thank M. Cariglia and M. Tonin for useful discussions.
This work is supported in part by the European Community's Human
Potential Programme under contract HPRN-CT-2000-00131 Quantum
Spacetime.



\end{document}